\documentclass[submission, PhysCodeb]{SciPost}

\hypersetup{
    colorlinks,
    linkcolor={red!50!black},
    citecolor={blue!50!black},
    urlcolor={blue!80!black}
}

\usepackage[bitstream-charter]{mathdesign}
\DeclareSymbolFont{usualmathcal}{OMS}{cmsy}{m}{n}
\DeclareSymbolFontAlphabet{\mathcal}{usualmathcal}

\usepackage{amssymb}
\usepackage{amsmath}
\usepackage{physics}
\usepackage{mathtools}
\usepackage{array}
\usepackage{ltablex}
\usepackage{caption}
\usepackage{multirow}
\usepackage{inputenc}[utf8]
\usepackage{graphicx}
\usepackage{subfig}
\usepackage[section]{placeins}

\usepackage{siunitx}    
\usepackage{booktabs}
\usepackage[referable]{threeparttablex} 
\usepackage{dcolumn}
\newcolumntype{Y}{D..{3}}

\usepackage[frozencache,cachedir=.]{minted}
\usepackage[most]{tcolorbox}
\tcbuselibrary{minted,skins}

\definecolor{pycodebg}{rgb}{0.95,0.95,0.95}
\definecolor{pyoutputbg}{rgb}{1.0,1.0,1.0}
\newtcblisting{pycode}{
  listing engine=minted,
  colback=pycodebg,
  colframe=black!25,
  listing only,
  minted language=py,
  minted options={linenos=false,texcl=false,fontsize=\small},
  left=1mm,
  right=1mm,
  top=1mm,
  bottom=1mm,
  breakable,
}
\newtcblisting{bashcode}{
  listing engine=minted,
  colback=pycodebg,
  colframe=black!25,
  listing only,
  minted language=bash,
  minted options={linenos=false,texcl=false,fontsize=\small},
  left=1mm,
  breakable,
}
\newtcblisting{exampleout}{
  listing engine=minted,
  colback=pyoutputbg,
  colframe=white,
  listing only,
  minted language=text,
  minted options={linenos=false,texcl=false,fontsize=\small},
  left=1mm,
  breakable,
}

\usepackage{verbatimbox} 

\begin{document}
\begin{center}{\Large \textbf{
Quantum MASALA: Quantum MAterialS Ab initio eLectronic-structure pAckage\\
}}\end{center}


\begin{center}
Shri Hari Soundararaj$^{\dagger \ddagger}$,
Agrim Sharma$^\dagger$ and
Manish Jain\textsuperscript{$\star$}
\end{center}

\begin{center}
Centre for Condensed Matter Theory, Department of Physics,\\
Indian Institute of Science, Bangalore 560012, India\\
${}^\star${\small\sf mjain@iisc.ac.in}
\end{center}

\begin{center}
\today
\end{center}


\def\thefootnote{$\dagger$}\footnotetext{These authors contributed equally to this work}\def\thefootnote{\arabic{footnote}}
\def\thefootnote{$\ddagger$}\footnotetext{Currently at Materials Science and Engineering, University of California Riverside, Riverside, CA 92521, USA}\def\thefootnote{\arabic{footnote}}
\section*{Abstract}
{\bf

We present \texttt{Quantum MASALA}, a compact package that implements different electronic-structure methods in Python using the plane-wave basis.
Within just 8100 lines of pure Python code, we have implemented Density Functional Theory (DFT), Time-dependent Density Functional Theory (TD-DFT) and the GW Method. The program can run across multiple processors and in Graphical Processing Units (GPU) with the help of easily-accessible Python libraries. With \texttt{Quantum ESPRESSO} and \texttt{BerkeleyGW} input interfaces implemented, it can also be used as a substitute for small and medium scale calculations, making it a perfect learning tool for {\em ab initio} methods. The package is aimed to provide a framework with its modular and simple code design to rapidly build and test new methods for first-principles calculation.
}

\vspace{10pt}
\noindent\rule{\textwidth}{1pt}
\tableofcontents\thispagestyle{fancy}
\noindent\rule{\textwidth}{1pt}
\vspace{10pt}

\section{Introduction}
Over the last few decades, {\em ab initio} electronic structure calculations have become a widespread methodology for studying various properties of molecules and materials. The exponential growth of computation power along with its accessibility has made first-principle calculations ubiquitous in condensed matter studies. Due to its enormous success, density functional theory (DFT)\cite{HK, KS} has become the standard technique for computational analysis of both molecular and periodic systems. The GW method\cite{Hedin65}, although computationally more expensive, has been proven to describe excitations in systems with high accuracy \cite{BGW2012}.

The development of robust, efficient and massively-parallel computer codes that implement such techniques has massively contributed to their widespread use. DFT Codes like \texttt{Quantum ESPRESSO}\cite{QE1, QE2}, \texttt{ABINIT}\cite{ABINIT1, ABINIT2}, \texttt{VASP}\cite{VASP_1996_PhysRevB.54.11169}, \texttt{PARSEC}\cite{PARSEC} and \texttt{Octopus}\cite{Octopus1, Octopus2} have been in active development for decades during which they evolved into feature-rich packages built to be efficient and massively scalable. Similarly, several packages such as \texttt{BerkeleyGW}\cite{BGW2012} and \texttt{YAMBO}\cite{YAMBO_2009} have a massively parallel implementation of the GW method which has been demonstrated to solve systems containing thousands of atoms.

While being efficient and scalable, the aforementioned packages are not best suited for quickly `prototyping' new ideas and algorithms. During the long development period of these packages, they have evolved into monoliths containing hundreds of thousands of lines of code, resulting in a steep learning curve for newcomers to the package. The general lack of flexibility/customizability in the low-level routines leads to much slower rate of development.

The recent developments in the field of data science have motivated the development of high-throughput screening of materials in order to accelerate the discovery of novel materials and structures\cite{aflow, aiida, fireworks, pymatgen, pygwbse}. The application of machine learning to electronic structure calculations is also an active field of research. With the rising interests in studying larger and more complex materials in search of novel properties, there is a need to formulate new methods/algorithms that improve the speed and efficiency of {\em ab initio} calculations, which is usually the performance bottleneck of such systems.

To address the need for a compact framework, we have designed \texttt{Quantum MASALA}, a Python \cite{CPython} package aimed to be a suite of electronic structure methods with a simple code design\footnote{\texttt{Quantum MASALA} GitHub repository: \url{https://github.com/qtm-iisc/Quantum MASALA}}. With less than 8100 lines of pure Python code\footnote{Excluding I/O interfaces with external codes}, \texttt{Quantum MASALA} implements the pseudopotential based DFT\cite{Ihm_1979}, time-dependent DFT\cite{Octopus1,Octopus2} and GW method\cite{HybertsenLouie86} for periodic solids. \texttt{Quantum MASALA} is designed with minimal required dependencies and supports multiple optional dependencies that enable MPI-parallelism and GPU-Acceleration, maximizing performance in any device. Written in one of the most popular programming languages that is known for its readable syntax, the package aims to provide a sandbox for implementing and testing new ideas quickly.

This article gives a complete description of the \texttt{Quantum MASALA} package. \texttt{Quantum MASALA} is a plane-wave basis based python package which implements different electronic-structure methods. We begin with a brief overview of the theory behind the implemented electronic structure methods which include, the standard plane-wave pseudopotential DFT method, time dependent DFT via a real-time propagation and the GW method. We describe the codebase, starting with a short discussion about runtime performance in Python. We discuss the general layout of the code which includes a brief description of the package's core modules and structures. The algorithms that are implemented in major calculation routines are described in detail. Finally, we demonstrate the capabilities of the package and provide an analysis of its performance and accuracy in comparison with popular packages. We conclude with a brief summary of the package's features and future development plans of the code.

\section{Theoretical Background}
\subsection{Density functional theory and the Kohn-Sham equations}
Density functional theory relates the ground-state electron density of a given system directly to the total energy. The Kohn-Sham formulation\cite{KS} of DFT maps the density of the many-body problem to an independent-particle system that is described by the Kohn-Sham (KS) Hamiltonian (in atomic units):
\begin{align}
\hat{H}_{\textrm{KS}}\ket{\psi_i} &= \bqty{-\frac{1}{2}\laplacian_i + V_{\textrm{aux}}(\mathbf{r}; n(\mathbf{r}))}\ket{\psi_i} \\
n(\mathbf{r}) &= \sum_i f_i\abs{\psi_i(\mathbf{r})}^2
\end{align}
where \(f_i\) denotes the occupation number of the state \(\ket{\psi_i}\).

The first term of \(\hat{H}_{\textrm{KS}}\) is the single-particle kinetic energy operator. The local potential \(V_{\textrm{aux}}(\mathbf{r}, n(\mathbf{r}))\) depends on the ground-state density \(n(\mathbf{r})\) and is given by the following:
\begin{equation}
V_{\textrm{aux}}(\mathbf{r}; n(\mathbf{r})) = V_{\textrm{ext}}(\mathbf{r}) + V_{\textrm{Hartree}}(\mathbf{r}; n(\mathbf{r})) + V_{\textrm{xc}}(\mathbf{r}; n(\mathbf{r}))
\end{equation}
where \(V_{ext}\) is the external potential, which in the case of the electronic structure problems is the Coulomb potential generated by the nuclei. The Hartree potential is the classical potential on an electron due to charge density and the exchange-correlation (XC) potentials represents the remaining many-body interaction. Note that \(V_{\textrm{xc}}\) can be spin-dependent based on the choice of the XC functional.

The density dependence of the potential in \(\hat{H}_{\textrm{KS}}\) requires the ground-state solution to be self-consistent with the potential in the Hamiltonian. Therefore, solving the KS system requires an self-consistent-field (SCF) iteration, which involves the following sequence of quantities to be evaluated:
\begin{align}
n_1 \rightarrow V_1 \rightarrow \qty{\ket{\psi_i}}_1 \rightarrow &n_{\textrm{out},1} \xRightarrow[\text{mix}]{n_{\textrm{out},1} \neq n_1} n_2 \\
n_2 \rightarrow V_2 \rightarrow \qty{\ket{\psi_i}}_2 \rightarrow &n_{\textrm{out},2} \xRightarrow[\text{mix}]{n_{\textrm{out},2} \neq n_2} n_3 \\
 &\vdots \\
n_N \rightarrow V_N \rightarrow \qty{\ket{\psi_i}}_N \rightarrow &n_{\textrm{out},N} \xRightarrow{n_{\textrm{out},N} = n_N} n_{N} \equiv n_{\textrm{KS}}
\end{align}
The number of steps required for achieving self-consistency can be reduced using fixed-point iteration algorithms, which involves a mixing of densities of input, output, and in some methods, values from previous iterations too.

\subsection{Wavefunctions in Periodic Solids and the Plane-Wave Basis}
In periodic solids, according to Bloch's Theorem, the wavefunctions of the KS Hamiltonian can be written as:
\begin{equation}
\ket{\psi_i} \rightarrow \ket{\psi_{\mathbf{k}}^i}; \braket{\mathbf{r}}{\psi_{\mathbf{k}}^i} = \psi_{\mathbf{k}}^i(\mathbf{r}) := e^{i\mathbf{k}\cdot \mathbf{r}}u_{\mathbf{k}}^i(\mathbf{r})
\end{equation}
where \(u_{\mathbf{k}}^i(\mathbf{r})\) has the periodicity of the crystal. The periodic function \(u_{\mathbf{k}}^i(\mathbf{r})\) can be expressed as a Fourier series, using the plane wave expansion
\begin{align}
\braket{\mathbf{r}}{\psi_{\mathbf{k}}^i} &= e^{i\mathbf{k}\cdot \mathbf{r}}\cdot\sum_{\mathbf{G}} u_{\mathbf{k}}^i(\mathbf{G}) e^{i\mathbf{G}\cdot \mathbf{r}} \\
&= \sum_{\mathbf{G}} u_{\mathbf{k}}^i(\mathbf{G}) \braket{\mathbf{r}}{\mathbf{G}+\mathbf{k}}; \braket{\mathbf{r}}{\mathbf{G} + \mathbf{k}} := e^{i(\mathbf{G} + \mathbf{k})\cdot \mathbf{r}}
\end{align}
where \(\mathbf{G}\) are the lattice points of the reciprocal lattice of the crystal.

The KS Hamiltonian in the plane wave (PW) basis takes the following form:
\begin{align}
\label{KSHamMat}
\mel{\mathbf{G}_m + \mathbf{k}}{\hat{H}_{\textrm{KS}}}{\mathbf{G}_n+\mathbf{k}} &= \mel{\mathbf{G}_{m + \mathbf{k}}}{\bqty{-\frac{1}{2}\laplacian + V_{\textrm{KS}}(\mathbf{r})}}{\mathbf{G}_n + \mathbf{k}}\\
H_{m,n}(\mathbf{k}) &= \frac{\abs{\mathbf{k} + \mathbf{G}_m}^2}{2}\delta_{m,n} + V_{\textrm{KS}}(\mathbf{G}_m - \mathbf{G}_n)
\end{align}
For solving the system numerically, the PW basis is truncated. The wave functions are expanded including only those plane waves whose energy is less than a energy cutoff, \(E_{\textrm{cut}}\) {\em i.e.} ($\abs{\mathbf{k} + \mathbf{G}_m}^2 / 2 \leq E_{\textrm{cut}}$). This implies that the PW expansion of \(V_{\textrm{KS}}\) must contain \(G\) within the energy cutoff \(4E_{\textrm{cut}}\) as $\abs{\mathbf{G}_m - \mathbf{G}_n} \leq \abs{\mathbf{G}_m} + \abs{\mathbf{G}_n} \leq 2\sqrt{2 E_{\textrm{cut}}}$

\subsection{Pseudopotentials}
Most chemical behaviour of atoms originate from the behaviour of the valence electrons which is influenced by their surrounding chemical environment. The core electrons, which remain unaffected under most situations, can be replaced with an effective `pseudo-potential' on the valence electrons that can accurately reproduce their states across a wide range of conditions. As pseudopotentials aim to replace the potential generated by the nucleus and core electrons, they must be spherically symmetric where each \((l,m)\) is treated individually. The `semi-local' form of a pseudopotential of an atom centred at origin is given below:
\begin{equation}
\hat{V}_{ps} = \sum_{l}\sum_{m=-l}^{l} \ket{l,m}V^l(r)\bra{l,m}
\end{equation}
This form can be simplified to a sum of projection operators, as given by Kleinman and Bylander\cite{KB}:
\begin{equation}
\hat{V}_{ps} = V_{loc}(\mathbf{r}) + \sum_{(l,m)} \frac{\ketbra{\Delta V_{ps}^l \varphi_{lm}}{\Delta V_{ps}^l \varphi_{lm}}}{\expval{\Delta V_{ps}^l}{\varphi_{lm}}};\; \Delta V_{ps}^l = V^l - V_{loc}
\end{equation}

\subsection{Time-dependent Density Functional Theory}
The Time-Dependent Density Functional Theory\cite{RG} extends DFT from describing ground-state properties to evaluating the real-time dynamics, enabling it to compute excited state properties and response to time-dependent perturbations.
The time-dependent Kohn-Sham (TDKS) equations take a form that is analogous to the time-dependent Schrodinger equation:
\begin{align}
i\pdv{t} \ket{\psi_i} &= \hat{H}_{\textrm{KS}}(t)\ket{\psi_i}\\
&= \bqty{-\frac{1}{2}\laplacian + V_{\textrm{KS}}(\mathbf{r}, t; n(\mathbf{r}, t))}\ket{\psi_i}\\
n(\mathbf{r}, t) &= \sum_i f_i\abs{\psi_i(\mathbf{r},t)}^2
\end{align}
The potential term \(V_{\textrm{KS}}(\mathbf{r}, t)\) here is analogous to the time-independent equations, containing a time-varying external potential \(V_{ext}(\mathbf{r}, t)\). For the XC Potentials in TDDFT, we limit ourselves to adiabatic approximations where \(V_{\textrm{xc}}(\mathbf{r},t;n(\mathbf{r}, t)) := V_{\textrm{xc}}(\mathbf{r}; n(\mathbf{r}, t))\)

The formal solution of the TDKS Equations is given by:
\begin{align}
\ket{\psi_i(t)} &= \mathcal{T}\exp{-i \int_0^t \dd{\tau} \hat{H}_{\textrm{KS}}(\tau)}\ket{\psi_i(t=0)} \\
&= \hat{U}_{\textrm{KS}}(t,0)\ket{\psi_i(t=0)}
\end{align}
where \(\mathcal{T}\) denotes the time-ordering operator. The time-evolution operator \(\hat{U}(t', t)\) can be split into smaller time-steps:
\begin{equation}
\hat{U}(t_N, t_0) = \hat{U}(t_N, t_{N-1})\cdot\hat{U}(t_{N-1}, t_{N-2})\cdots \hat{U}(t_2, t_1)\cdot \hat{U}(t_1, t_0);\; t_n = t_0 + n\Delta t
\end{equation}

TDDFT can be used to compute the optical absorption spectra by propagating the TDKS equations. By perturbing the ground-state system with a delta electric field along, lets say x-axis, \(v_{pert}(\mathbf{r}, t) = -k_0\delta(t) \cdot(\mathbf{r}\cdot\hat{x})\), the system evolves in time with a density response \(\delta n(\mathbf{r}, t) = n(\mathbf{r}, t) - n_0(\mathbf{r})\). For small perturbation \(k_0 << 1\), the response is expected to be linear and dipolar, allowing us to compute the dynamical polarizability from the dipole response
\begin{equation}
\alpha_{x\nu}(\omega) = - \frac{1}{k_0} \int \dd[3]r \delta n(\mathbf{r}, \omega) r_\nu;\;\ \nu=x,y,z
\end{equation}
The absorption cross-section is proportional to the imaginary part of \(\alpha\) averaged over the three spatial directions.
\begin{equation}
\sigma(\omega) = \frac{4\pi\omega}{c}\cdot \mathfrak{Im} \frac{1}{3} \sum_{\nu} \alpha_{\nu\nu}(\omega)
\end{equation}

\subsection{GW Approximation}

Kohn-Sham DFT yields the correct ground state electron density, but the corresponding non-interacting single-particle orbitals need not have any direct physical interpretation\cite{Hedin65, ShamSchluter, PerdewLevy}. Electron addition and removal energies are quasiparticle energies and its calculation requires many-body effects to be taken into account\cite{Hedin65, HybertsenLouie86, Aryasetiawan_1998}. This can be done using Hedin's GW formalism \cite{Hedin65}, where one directly calculates the quasiparticle energies as poles of the single particle Green's function. Within this formalism, the exact many-body self-energy is expressed as a perturbation series in terms of dynamically screened Coulomb interaction.

While in principle, to calculate the poles of the single-particle Greens function one needs to solve the Dyson equation, often, the quasiparticle energies are computed as a single shot correction to the DFT energy eigenvalues. Within this approach the Greens function ($\textrm{G}_0$) and screened Coulomb interaction ($\textrm{W}_0$) are constructed from DFT orbitals and energy eigenvalues. For a large variety of materials, the first iteration is good enough for calculating the quasiparticle energy spectrum\cite{PRB_Holm_Barth}. This method of doing a single GW iteration is known as the \textit{G\textsubscript{0}W\textsubscript{0} method}. 

Written explicitly, we are interested in solving
\begin{equation}\label{eqn:qp_schr_eqn}
    \left[ -\frac{1}{2}\hat{\nabla}^2+\hat{V}_{\textrm{ext}}+\hat{V}_{\textrm{Hartree}}+\hat{\Sigma}(E_{n\bf k}) \right]\,\psi_{n\bf k}^{\text{QP}} = E_{n\bf k}\,\psi_{n\bf k}^{\text{QP}}
\end{equation}
where $\hat{\Sigma}$ is the G\textsubscript{0}W\textsubscript{0} dynamical self energy operator which is non-local, and non-Hermitian in general\cite{HybertsenLouie86}. As a result, the quasiparticle energy eigenvalues $E_{n\bf k}$ are complex - their imaginary parts being inversely related to quasiparticle lifetimes. While calculating the quasiparticle energy spectrum, $E_{n\textbf{k}}^{\textrm{QP}} = \textrm{Re}\left(E_{n\textbf{k}}\right)$ for quasiparticles with long lifetimes, it is common practice to solve the real part of eqn. \ref{eqn:qp_schr_eqn}, while evaluating the self energy operator $\hat{\Sigma}(E)$ at $E_{n\textbf{k}}^{\textrm{QP}}$, instead of $E_{n\textbf{k}}$. Further, in a typical calculation, the DFT eigenvectors are sufficiently close to the quasiparticle eigenvectors \cite{BGW2012} for the following approximation to be used:
\begin{equation}
    E_{n\bf k}^{\text{QP}} = E_{n\bf k}^{\text{DFT}} + \bra{\psi_{n\bf k}^{\text{DFT}}}\left(\hat{\Sigma}\left(E_{n\bf k}^{\textrm{QP}}\right)-V_{\textrm{xc}}\right)\ket{\psi_{n\bf k}^{\text{DFT}}}
\end{equation}
The $G_0W_0$ self-energy $\Sigma$ is defined as\cite{Hedin65}:
\begin{equation}
    \Sigma\left(\textbf{r}, \textbf{r}';t,t'\right) = i\, G_0\left(\textbf{r}, \textbf{r}';t,t'\right)\, W_0\left(\textbf{r}, \textbf{r}';t\!+\!0^+,t'\right)
\end{equation}
where $G_0(\textbf{r}, \textbf{r}';t,t')$ is the single particle Green's function 
\begin{equation}
G_0\left(\textbf{r}, \textbf{r}'; t, t'\right) = -i\bra{N}T\left[\hat{\psi}^\dagger\left(\textbf{r}', t'\right)\,\hat{\psi}\left(\textbf{r}, t\right)\right]\ket{N}
\end{equation}
and $W_0(\textbf{r}, \textbf{r}';t,t')$ is screened Coulomb interaction 
\begin{equation}
    W_0\left(\textbf{r}, \textbf{r}'; t, t'\right)=\int \epsilon^{-1}\left(\textbf{r},\textbf{r}''; t, t'\right)\, v\left(\textbf{r}'',\textbf{r}'\right) \,\textrm{d}\textbf{r}''
\end{equation}
In time translation invariant systems, the self energy effectively becomes a function of $t-t'$ instead of $t$ and $t'$. The same holds for $G_0$, $W_0$, and $\epsilon$.

In the rest of the section we will use Rydberg units. In the plane wave basis, the random phase approximation (RPA) dielectric function is:
\begin{equation}
\epsilon_{\textbf{GG'}}{\left({\textbf q}\;\!;\omega\right)}=
\delta_{\textbf{GG'}}\,{-}\,v{\left({\textbf q}{+}{\textbf G}\right)}\,
P_{\textbf{GG'}}{\left({\textbf q}\;\!;\omega\right)}
\label{epsilon}
\end{equation}
where $\textbf G$ and $\textbf G'$ are reciprocal lattice vectors, $\textbf q$ lies in the first Brillouin zone, $v(\textbf q+\textbf G)$ is bare Coulomb potential:
\begin{equation}
    v(\textbf{q} + \textbf{G}) = \frac{8\pi}{\left|\textbf q + \textbf G\right|^2}
\end{equation}
and $P$ is irreducible polarizability, for which the Adler-Wiser expression simplifies to\cite{BGW2012}:
\begin{equation}
P_{\textbf{GG'}}{\left({\textbf q}\;\!;\omega=0\right)}=
\,\,{}\sum_{n}^{\textrm{occ}}\sum_{n'}^{\textrm{emp}}\sum_{{\textbf k}}
\frac{
\bra{n'\textbf k}e^{-i({\textbf q}+{\textbf G})\cdot{\textbf r}}\ket{n{\textbf k}{+}{\textbf q}}
\bra{n{\textbf k}{+}{\textbf q}}e^{i({\textbf q}+{\textbf G'})\cdot{\textbf r}}\ket{n'\textbf k}
}{E_{n{\textbf k}{+}{\textbf q}}\,{-}\,E_{n'{\textbf k}}}.
\label{eqn_static_xi}
\end{equation}


In the GW formalism, the real part of self-energy as defined above naturally splits into two parts: the product of the real parts of $G_0$ and $W_0$ gives the screened exchange (SX) term and the product of their imaginary parts gives the Coulomb-hole (CH) term. \cite{ZhangLouie89} 
\begin{equation}
    \Re\left(\Sigma\right) = \Sigma_\textrm{SX} + \Sigma_\textrm{CH}
\end{equation}


Under the static COHSEX approximation proposed by Hedin\cite{Hedin65}, the self-energy matrices can be expressed as:
\begin{align}
\label{cohsex-sx}
\left<n{\textbf k}\right|\Sigma_{\textrm{SX}}{\left(E=0\right)}\left|n'{\textbf k}\right>=
-\sum_{n''}^{\textrm{occ}}\sum_{\textbf{qGG'}}&
\left<n{\textbf k}\right|e^{i({\textbf q}+{\textbf G})\cdot{\textbf r}}\left|n''{\textbf k}{-}{\textbf q}\right>
\left<n''{\textbf k}{-}{\textbf q}\right|e^{-i({\textbf q}+{\textbf G}')\cdot{\textbf r}'}\left|n'{\textbf k}\right>
 \nonumber\\
&\text{ }\;\;\times
\epsilon_{\textbf{GG'}}^{-1}{\left({\textbf q}\;\!;0\right)}
v{\left({\textbf q}{+}{\textbf G}'\right)}
\end{align}

\begin{align}
\label{cohsex-ch-sum}
\left<n{\textbf k}\right|\Sigma_{\textrm{CH}}{\left(E=0\right)}\left|n'{\textbf k}\right>&=
\frac{1}{2}\sum_{n''}\sum_{\textbf{qGG'}}
\left<n{\textbf k}\right|e^{i({\textbf q}+{\textbf G})\cdot{\textbf r}}\left|n''{\textbf k}{-}{\textbf q}\right>
\left<n''{\textbf k}{-}{\textbf q}\right|e^{-i({\textbf q}+{\textbf G}')\cdot{\textbf r}'}\left|n'{\textbf k}\right>
\nonumber \\
&\text{\hspace{6em}}\times
\left[\epsilon_{\textbf{GG'}}^{-1}{\left({\textbf q}\;\!;0\right)}
\,{-}\,\delta_{\textbf{GG'}}\right]
v{\left({\textbf q}{+}{\textbf G}'\right)}\\
&=\frac{1}{2}\sum_{\textbf{qGG'}}
\left<n{\textbf k}\right|e^{i({\textbf G}-{\textbf G}')\cdot{\textbf r}}\left|n'{\textbf k}\right>
\left[\epsilon_{\textbf{GG'}}^{-1}{\left({\textbf q}\;\!;0\right)}
\,{-}\,\delta_{\textbf{GG'}}\right]
v{\left({\textbf q}{+}{\textbf G}'\right)} \label{cohsex-exact-ch}
\end{align}
We note that the completeness relation of mutually orthogonal orbitals has been used in (eq.\ref{cohsex-exact-ch}), to skip the expensive summation over all bands.
Also note that conventionally, the self energy is expanded as the sum of a Hartree-Fock exchange term and a (Coulomb) correlation term:
\begin{equation}
    \Sigma = \Sigma_\textrm{X} + \Sigma_\textrm{Corr}
\end{equation}
The Hartree-Fock exchange matrix elements can be calculated as follows:
\begin{align}
\label{hf-x}
\left<n{\textbf k}\right|\Sigma_{\textrm{X}}\left|n'{\textbf k}\right>=
-\sum_{n''}^{\textrm{occ}}\sum_{\textbf{qGG'}}&
\left<n{\textbf k}\right|e^{i({\textbf q}+{\textbf G})\cdot{\textbf r}}\left|n''{\textbf k}{-}{\textbf q}\right>
\left<n''{\textbf k}{-}{\textbf q}\right|e^{-i({\textbf q}+{\textbf G}')\cdot{\textbf r}'}\left|n'{\textbf k}\right> 
 \nonumber\\
&\text{ }\;\;\times
\delta_{\textbf{GG'}}
v{\left({\textbf q}{+}\textbf{G'}\right)}
\end{align}
The correlation term captures the effect of screening due to other electrons. We can identify the real part of the correlation term to be $\textrm{Re}\left(\Sigma_\textrm{Corr}\right)=\left(\Sigma_\textrm{SX}-\Sigma_\textrm{X}\right)+\Sigma_\textrm{CH}$.

Beyond Static-COHSEX, dynamical self-energy needs to be solved self-consistently, as shown below. 
\begin{align}
    E_{n\textbf{k}}^{\textrm{QP}} &= \textrm{Re}\left(E_{n\textbf{k}}\right)\\    &=\textrm{Re}\left(\Sigma_{n\textbf{k}}\left(E_{n\textbf{k}}^{\textrm{QP}}\right)\right) + E_{n\bf k}^{\text{DFT}} - V_{\textrm{xc},n\bf k}^{\text{DFT}}
\end{align}
The Hybertsen-Louie plasmon pole model \cite{HybertsenLouie86} assumes that contribution from a single plasmon dominates the pole structure of the inverse dielectric function. Using Kramers-Kronig relations and a generalized f-sum rule \cite{HybertsenLouie86}, one can get rid of all free parameters in the single plasmon dielectric function \cite{ZhangLouie89}, and the static dielectric function value can be used to generate values at all frequencies. With this approximation, we obtain the following self-energy matrix elements:
\begin{align}
\label{gpp-sx}
\left<n{\bf k}\right|\Sigma_{\textrm{SX}}{\left(E\right)}\left|n'{\bf k}\right>=
-\sum_{n''}^{\textrm{occ}}\sum_{{\bf qGG}'}
\left<n{\bf k}\right|e^{i({\bf q}+{\bf G})\cdot{\bf r}}\left|n''{\bf k}{-}{\bf q}\right>
\left<n''{\bf k}{-}{\bf q}\right|e^{-i({\bf q}+{\bf G}')\cdot{\bf r}'}\left|n'{\bf k}\right>\\
\nonumber
\times
\left[\delta_{{\bf GG}'}+\frac{\Omega^2_{{\bf GG}'}{\left({\bf q}\;\!\right)}
\left(1\,{-}\,i\tan\phi_{{\bf GG}'}{\left({\bf q}\;\!\right)}\right)}
{\left(E\,{-}\,E_{n''{\bf k}{-}{\bf q}}\right)^2\!{-}\:
\tilde{\omega}^2_{{\bf GG}'}{\left({\bf q}\;\!\right)}}\right]
v{\left({\bf q}{+}{\bf G}'\right)}
\end{align}
\begin{align}
\label{gpp-ch}
\left<n{\bf k}\right|\Sigma_{\textrm{CH}}{\left(E\right)}\left|n'{\bf k}\right>=
\frac{1}{2}\sum_{n''}\sum_{{\bf qGG}'}
\left<n{\bf k}\right|e^{i({\bf q}+{\bf G})\cdot{\bf r}}\left|n''{\bf k}{-}{\bf q}\right>
\left<n''{\bf k}{-}{\bf q}\right|e^{-i({\bf q}+{\bf G}')\cdot{\bf r}'}\left|n'{\bf k}\right>\\
\nonumber
\times\,\frac{\Omega^2_{{\bf GG}'}{\left({\bf q}\;\!\right)}
\left(1\,{-}\,i\tan\phi_{{\bf GG}'}{\left({\bf q}\;\!\right)}\right)}
{\tilde{\omega}_{{\bf GG}'}{\left({\bf q}\;\!\right)}
\left(E\,{-}\,E_{n''{\bf k}{-}{\bf q}}{-}\,
\tilde{\omega}_{{\bf GG}'}{\left({\bf q}\;\!\right)}\right)}
\;v{\left({\bf q}{+}{\bf G}'\right)}
\end{align}
where the plasmon mode parameters are defined as follows:\\
The effective bare plasma frequency:
\begin{equation}
\label{capitalomega}
\Omega^2_{{\bf GG}'}{\left({\bf q}\;\!\right)}=\omega_{\textrm{p}}^2\,\,\frac{{\left({\bf q}{+}{\bf G}\right)}{\cdot}{\left({\bf q}{+}{\bf G}'\right)}}{\left|{\bf q}{+}{\bf G}\right|^2}\,\,\frac{\rho{\left({\bf G}{-}{\bf G}'\right)}}{\rho{\left({\bf 0}\right)}}
\end{equation}
the GPP mode frequency:
\begin{equation}
\label{tildeomega}
\tilde{\omega}^2_{{\bf GG}'}{\left({\bf q}\;\!\right)}=\frac{\left| \lambda_{{\bf GG}'}{\left({\bf q}\;\!\right)} \right|} {\cos\phi_{{\bf GG}'}{\left({\bf q}\;\!\right)}}
\end{equation}
and the renormalized bare plasmon frequency:
\begin{equation}
\label{phi}
\left| \lambda_{{\bf GG}'}{\left({\bf q}\;\!\right)} \right| e^{i\phi_{{\bf GG}'}{\left({\bf q}\;\!\right)}}=\frac{\Omega^2_{{\bf GG}'}{\left({\bf q}\;\!\right)}}{\delta_{{\bf G}{\bf G}'}{-}\epsilon_{{\bf G}{\bf G}'}^{-1}({\bf q};0)}
\end{equation}
where, $\rho$ denotes the electron charge density and $\omega_{\textrm{p}}^2=4 \pi \rho({\bf 0}) e^2 / m$ denotes the classical plasma frequency.

Divergent cases arising from the above equations are handled as per the prescription given in \texttt{BerkeleyGW}\cite{BGW2012}. Due to the $n''$-dependence of $\epsilon_{\textrm{HLPP}}$, it is not possible to get rid of the $n''$ summation over all orbitals in $\Sigma_{\textrm{CH}}$ calculation. However, several methods have been devised to reduce the number of empty orbitals required to make this sum converge. Static remainder approach discussed in \cite{StaticRemainder2013} is one such scheme,
\begin{align}
\label{static-remainder}
\left<n{\bf k}\right| \Sigma_{\textrm{CH}}^\infty{\left(E\right)}\left|n'{\bf k}\right>= 
& \left<n{\bf k}\right|\Sigma_{\textrm{CH}}^N{\left(E\right)}\left|n'{\bf k}\right> 
\\ \nonumber
&+\frac{1}{2}
\bigg(\left<n{\bf k}\right|\Sigma_{\textrm{CH}}^{\infty}{\left(E=0\right)}\left|n'{\bf k}\right>-\left<n{\bf k}\right|\Sigma_{\textrm{CH}}^{N}{\left(E=0\right)}\left|n'{\bf k}\right>\bigg)
\end{align}
where $N$ or $\infty$ denote the number of empty orbitals included in $n''$ summation and `$E=0$' denotes static-COHSEX self-energies.

Dynamic self-energy calculated with plasmon-pole method depend on energy parameter $E$. In order to avoid a series of self-consistent iterations to compute $E_{n\textbf{k}}^{\textrm{QP}}$, we compute $\Sigma_\textrm{HLPP}$ for two different sets of energies and then find $E_{n\textbf{k}}^{\textrm{QP}}$ using the secant method:
\begin{equation}
    E^{\textrm{QP}}_{n{\bf k}} = E^{0}_{n{\bf k}} + \frac{d\Sigma/dE}{1-d\Sigma/dE}\left(E^{0}_{n{\bf k}} - E^{\textrm{MF}}_{n{\bf k}} \right)
\end{equation}
where $E^{0}_{n{\bf k}} = \Sigma_{\textrm{HLPP}}\big(E^{\textrm{MF}}_{n{\bf k}}\big)$ and $d\Sigma/dE$ is calculated using a forward finite difference scheme.

\section{Writing performant code in Python: Libraries used}
Python, as a programming language, is considered to be slow in comparison to compiled languages like C, C++, FORTRAN, etc.
The reference implementation, CPython\cite{CPython}, is not as fast as other interpreted languages like MATLAB, Julia, R, etc. due to a lack of optimizations in its bytecode interpreter.
High-throughput in Python is usually achieved by libraries and packages that have their core implementations written in `faster' languages, usually C.
The data structures and routines implemented can be accessed from Python via defined interfaces.
For a Python program to be `efficient' in comparison to its analogues written in a compiled program, it is important to maximize the amount of processing that can be performed in each call to optimised routines implemented in these libraries.

There are alternative strategies for speeding up Python. Some of the relevant ones are listed below:
\begin{itemize}
\item PyPy\cite{PyPy} is an alternative implementation of Python that implements Just-In-Time (JIT) Compilation to speed up execution.
\item Numba\cite{Numba} is a third-party package that brings JIT Compilation to CPython.
\item Cython\cite{Cython} extends Python with C-like syntax and compilation support that results in performance similar to a C/C++ code.
\end{itemize}
We have refrained from using the aforementioned optimization tools in order to maximize compatibility and reduce dependencies, but we encourage developers to try them when extending our package for their needs wherever applicable.

\texttt{Quantum MASALA} uses NumPy\cite{NumPy} for performing all linear algebra operations. The library provides support for multi-dimensional array of homogeneous data along with an extensive set of routines implementing mathematical operations and linear algebra algorithms. It is easy-to-use and interoperable across most scientific computing packages in Python. The library calls optimized BLAS and LAPACK functions for linear algebra operations, allowing access to near C/FORTRAN speeds in Python.

Analogous to NumPy, CuPy\cite{CuPy} is a Python library that contains implementations of a subset of NumPy operations to run in an NVIDIA GPU. It provides a NumPy-like interface to Arrays in GPU Memory with nearly identical function calls to operate on them. This enables \texttt{Quantum MASALA} to run computationally intensive part of a calculation like FFT's, diagonalization, etc. in a GPU. As a part of its design, \texttt{Quantum MASALA} reuses a lot of the CPU code for its GPU implementation, which is possible due to NumPy's interoperability with CuPy.

\texttt{Quantum MASALA} can also use FFT Libraries like FFTW3 (via \texttt{pyFFTW}\cite{pyFFTW}) and MKL (via \texttt{mkl\_fft}\cite{mklfft}) if already installed. If not installed, the program automatically falls back to the slower FFT routines implemented in SciPy\cite{SciPy}.

\section{Usage: Core Objects of \texttt{Quantum MASALA}}


\subsection{Installation}
The code for \texttt{Quantum MASALA} is available as a git repository at \url{https://github.com/qtm-iisc/QuantumMASALA}. The code supports Python version 3.9 and beyond. 

These installation instructions require \texttt{conda}. If you do not have a functioning \texttt{conda} installation, we recommend installing it by following these quick-install instructions:
\url{https://docs.conda.io/projects/miniconda/en/latest/#quick-command-line-install}

\subsubsection{Installing on Mac}

To create the necessary environment, execute the following command:
\begin{bashcode}
conda create -n qtm python=3.12 numpy "libblas=*=*accelerate" \
      pyfftw pylibxc mpi4py scipy cython
\end{bashcode}

\textbf{Note:} For older versions of Python, it may be necessary to install numpy separately with Apple's Accelerate framework. With newer versions of numpy (> 2.0), the wheels are automatically built with Apple's Accelerate framework (Veclib).

Activate the environment:
\begin{bashcode}
conda activate qtm
\end{bashcode}

To test the numpy and scipy installation, install the following packages:
\begin{bashcode}
conda install -c conda-forge pytest hypothesis meson scipy-tests \
      pooch pyyaml
\end{bashcode}

\textbf{Note:} Please verify that the \texttt{python} command points to the environment's Python installation by running \texttt{which python}. If it does not point to the correct Python installation, we recommend specifying the full \texttt{python} path (i.e., \texttt{\$CONDA\_PREFIX/bin/python}) for the following commands.

To test numpy and scipy, use the following commands:
\begin{bashcode}
python
import numpy
numpy.show_config()
numpy.test()
import scipy
scipy.show_config()
scipy.test()
\end{bashcode}

For optimal performance, we recommend setting the relevant \texttt{...\_NUM\_THREADS} environment variables to 1:
\begin{bashcode}
conda env config vars set OMP_NUM_THREADS=1
conda env config vars set VECLIB_MAXIMUM_THREADS=1
conda env config vars set OPENBLAS_NUM_THREADS=1
\end{bashcode}

Reactivate the environment for the changes to take effect:
\begin{bashcode}
conda deactivate
conda activate qtm
\end{bashcode}

Inside the \texttt{QuantumMASALA} root directory, execute the following command to complete the installation:
\begin{bashcode}
python -m pip install -e .
\end{bashcode}

Test the installation by running the following example (replace \texttt{2} with the number of cores):
\begin{bashcode}
cd examples/dft-si
mpirun -np 2 python si_scf.py
\end{bashcode}

To perform a complete test, use \texttt{pytest} from the main directory which contains the tests folder:
\begin{bashcode}
python -m pip install pytest
pytest
\end{bashcode}

\subsubsection{Installing on Linux}

To create the environment, execute the following commands:
\begin{bashcode}
conda create -n qtm python=3.11
conda activate qtm
conda install -c conda-forge pyfftw pylibxc mpi4py
\end{bashcode}

If you are working on an Intel system, we recommend using \texttt{mkl\_fft} for optimal performance:
\begin{bashcode}
conda install -c conda-forge mkl_fft
\end{bashcode}

For optimal performance, we recommend setting the \texttt{OMP\_NUM\_THREADS} environment variable to 1:
\begin{bashcode}
conda env config vars set OMP_NUM_THREADS=1
\end{bashcode}

Reactivate the environment for the changes to take effect:
\begin{bashcode}
conda deactivate
conda activate qtm
\end{bashcode}

Inside the \texttt{QuantumMASALA} root directory, execute the following command to complete the installation:
\begin{bashcode}
python -m pip install -e .
\end{bashcode}

\textbf{Note:} Please verify that the \texttt{python} command points to the environment's Python installation by running \texttt{which python}. If it does not point to the correct Python installation, we recommend specifying the full \texttt{python} path (i.e., \texttt{\$CONDA\_PREFIX/bin/python}) for the following commands.

Test the installation by running the following example (replace \texttt{2} with the number of cores):
\begin{bashcode}
cd examples/dft-si
mpirun -np 2 python si_scf.py
\end{bashcode}

To perform a complete test, use \texttt{pytest} from the main directory which contains the tests folder:
\begin{bashcode}
python -m pip install pytest
pytest
\end{bashcode}


\subsubsection{Optional Dependencies}
The user may choose to install with optional libraries for added features. 
GPU support is provided by CuPy\cite{CuPy}. The instructions for installing \texttt{CuPy} are available at: \url{https://docs.cupy.dev/en/stable/install.html}. If a \texttt{CUDA} driver is already installed, the following command can be used to install both \texttt{CUDA-Toolkit} and \texttt{CuPy} from \texttt{conda-forge}:
\begin{bashcode}
conda install -c conda-forge cupy
\end{bashcode}
The Python package \texttt{primme} provides interface to \texttt{PRIMME}\cite{PRIMME}, a C library for iterative solution of eigenvalue problems. 
\begin{bashcode}
python -m pip install primme
\end{bashcode}

\subsubsection{Compiling the Documentation}

We use Sphinx to compile documentation from the docstrings. The documentation is available on 'Read the Docs' at \url{https://quantummasala.readthedocs.io/en/latest/}, and can also be compiled locally.

To compile the documentation locally using Sphinx, install the following packages:
\begin{bashcode}
python -m pip install sphinx sphinx-rtd-theme sphinx-math-dollar
\end{bashcode}

Compile the reStructuredText documentation by running the following from the root directory:
\begin{bashcode}
sphinx-apidoc -e -E --ext-autodoc --ext-mathjax -o docs/ -d 2 src/qtm/
\end{bashcode}

Build the documentation in HTML format by running the following from the root directory:
\begin{bashcode}
sphinx-build -b html ./docs ./docs/build
\end{bashcode}

\subsection{Defining a Crystal in Quantum MASALA}
In \texttt{Quantum MASALA}, the lattice of translations is represented by the \texttt{Lattice} class.
This class contains the primitive translation vectors and provides methods for coordinate transformation.
The \texttt{RealLattice} and the \texttt{ReciLattice} classes extend the \texttt{Lattice} class with lattice parameter `$a$' which is \texttt{alat} for \texttt{RealLattice} and \texttt{tpiba} ($2\pi / a$) for \texttt{ReciLattice}.
The two subclasses also provide methods to instantiate from instances of their duals.
Some usage examples are highlighted below:
\begin{pycode}
import numpy as np

from qtm.lattice import RealLattice, ReciLattice
from qtm.constants import ANGSTROM

# Defining a BCC Lattice
alat = 5.1 * ANGSTROM  # Lattice parameter 'a'
# Lattice vectors
latvec_alat = 0.5 * np.array([
    [ 1,  1,  1],
    [-1,  1,  1],
    [-1, -1,  1]
])

# Creating RealLattice instance representing BCC lattice
reallat = RealLattice.from_alat(alat, *latvec_alat)
# Creating ReciLattice instance from reallat
recilat = ReciLattice.from_reallat(reallat)

# Printing axes
print(reallat.alat, reallat.axes_alat)
## 9.637603235591428 ([0.5, 0.5, 0.5], [-0.5, 0.5, 0.5], [-0.5, -0.5, 0.5])

print(recilat.tpiba, recilat.axes_tpiba)
## 0.6519447993019614 ([1.0, 0.0, 1.0], [-1.0, 1.0, 0.0], [0.0, -1.0, 1.0])

# Input array of vectors in crystal coordinates
vec_cryst = np.array([
    [1, 0, 0],
    [0, 1, 0],
    [1, 1, 0],
    [0, 1, 1]
])

# Coordinate Transforms
vec_cart = reallat.cryst2cart(vec_cryst, axis=1)
print(vec_cart)
## [[ 4.81880162  4.81880162  4.81880162]
##  [-4.81880162  4.81880162  4.81880162]
##  [ 0.          9.63760324  9.63760324]
##  [-9.63760324  0.          9.63760324]]

vec_alat = reallat.cryst2alat(vec_cryst, axis=1)
print(vec_alat)
## [[ 0.5  0.5  0.5]
##  [-0.5  0.5  0.5]
##  [ 0.   1.   1. ]
##  [-1.   0.   1. ]]

# NOTE: As per convention, the first dimension of vector lists
# must correspond to the components, resulting in a shape of
# form (3, ...)
len_alat = reallat.norm(vec_cryst.T, coords='cryst')
print(len_alat / reallat.alat)
## [0.8660254  0.8660254  1.41421356 1.41421356]
\end{pycode}

A crystal can be fully specified by the lattice and the atoms in the unit cell. Similarly, the \texttt{Crystal} class in \texttt{Quantum MASALA} is defined by:
\begin{enumerate}
    \item A \texttt{RealLattice} instance describing the lattice of the crystal
    \item A sequence of \texttt{AtomBasis} instances describing the crystal's basis atoms
\end{enumerate}
The \texttt{AtomBasis} class represent a collection of atoms in the unit cell belonging to one atomic species/type and describes the position of the atoms in the crystal's unit cell along with the species' name, mass, and optionally its pseudopotential.
Note that for DFT calculations, the pseudopotential is a required argument. Currently \texttt{Quantum MASALA} supports the UPF v2 format only via the \texttt{UPFv2Data} container.

\begin{pycode}
""" Generating Crystal instance representing Iron """

from qtm.lattice import RealLattice
from qtm.crystal import BasisAtoms, Crystal
from qtm.pseudo import UPFv2Data
from qtm.constants import BOHR

# Defining the lattice
reallat = RealLattice.from_alat(alat=5.1070 * BOHR,
                                a1=[ 0.5,  0.5,  0.5],
                                a2=[-0.5,  0.5,  0.5],
                                a3=[-0.5, -0.5,  0.5])

# Reading pseudopotential data
fe_oncv = UPFv2Data.from_file('../examples/dft-fe/Fe_ONCV_PBE-1.2.upf')

# Specifying the basis atoms
fe_atoms = BasisAtoms.from_cart('fe', fe_oncv, 55.487, reallat,# 
           [0.0, 0.0, 0.0])

# Constructing the Crystal instance
crystal = Crystal(reallat, [fe_atoms, ])

# Printing crystal information
print(str(crystal))
\end{pycode}
\paragraph{Output}
\begin{verbatim}
Lattice parameter 'alat' : 5.10700  a.u.
Unit cell volume         : 66.59898  (a.u.)^3
Number of atoms/cell     : 1
Number of atomic types   : 1
Number of electrons      : 16

Crystal Axes: coordinates in units of 'alat' (5.10700 a.u.)
    a(1) = ( 0.50000,  0.50000,  0.50000)
    a(2) = (-0.50000,  0.50000,  0.50000)
    a(3) = (-0.50000, -0.50000,  0.50000)

Reciprocal Axes: coordinates in units of 'tpiba' (1.23031 (a.u.)^-1)
    b(1) = ( 1.00000,  0.00000,  1.00000)
    b(2) = (-1.00000,  1.00000,  0.00000)
    b(3) = ( 0.00000, -1.00000,  1.00000)


Atom Species #1
    Label   : fe
    Mass    : 55.49
    Valence : 16.00
    Pseudpot: Fe_ONCV_PBE-1.2.upf
              MD5: 0a0a3b4237ca738a29386c4fd1f8ac5e
    Coordinates (in units of alat)
          1 - ( 0.00000,  0.00000,  0.00000)

\end{verbatim}

\subsection{Bases in Plane-Wave Codes: Real-Space and G-Space}
Contrary to the name, plane-wave codes perform calculations in both reciprocal (Fourier) space and real space.
Although the former gives a compact representation of plane-waves, the latter is more effective for evaluating certain operations.
For instance, a local potential has a simple operator representation in real-space, where it forms a diagonal matrix.
But, the same in reciprocal basis would yield a dense matrix due to convolution.
\begin{align*}
\hat{V} &\coloneqq \int_{\Omega} \dd{\mathbf{r}} V(\mathbf{r}) \dyad{\mathbf{r}} \\
\mel**{\mathbf{G}}{\hat{V}}{\mathbf{G'}} &= \int_{\Omega} \dd{\mathbf{r}} V(\mathbf{r}) \braket{\mathbf{G}}{\mathbf{r}} \braket{\mathbf{r}}{\mathbf{G'}} \\
&= \int_{\Omega} \dd{\mathbf{r}} V(\mathbf{r}) \exp{-i \mathbf{G}\cdot \mathbf{r}} \exp{i \mathbf{G'}\cdot \mathbf{r}} \\
&= \sum_{\mathbf{G''}} \int_{\Omega}\dd{\mathbf{r}} V(\mathbf{G''}) \exp{i \mathbf{G''}\cdot \mathbf{r}} \exp{-i(\mathbf{G} - \mathbf{G'})\cdot \mathbf{r}} \\
&= \sum_{\mathbf{G''}} V(\mathbf{G''}) \delta_{\mathbf{G} - \mathbf{G'}, \mathbf{G''}}
\end{align*}
Thanks to Fast Fourier Transforms, it is easier and more efficient to implement the operator in the following way:

\begin{enumerate}
\item Inverse Fourier Transform the plane-wave wave functions
\begin{equation}
\psi_{i, \mathbf{k}}(\mathbf{G}) \xrightarrow{IFFT} \psi_{i, \mathbf{k}}(\mathbf{r})
\end{equation}
\item Apply the real-space potential by simple element-wise multiplication
\begin{equation}
\hat{V}\ket{\psi_{i,\mathbf{k}}} \equiv \int_{\Omega} \dd{\mathbf{r}} \dyad{\mathbf{r}} V(\mathbf{r})\psi_{i, \mathbf{k}}(\mathbf{r})
\end{equation}
\item Forward Fourier Transform the result
\begin{equation}
\bqty{\hat{V}\psi_{i,\mathbf{k}}}(\mathbf{r}) \xrightarrow{FFT} \bqty{\hat{V}\psi_{i, \mathbf{k}}}(\mathbf{G})
\end{equation}
\end{enumerate}

Similarly, certain quantities such as the kinetic energy operator, the Hartree potential, etc. are simpler in reciprocal space than in real-space. So, they can be evaluated in reciprocal space as is and can be transform to real-space if needed.
\begin{align}
\mel{\mathbf{q}}{\hat{K}}{\mathbf{q'}} &= \mel{\mathbf{q}}{-\frac{1}{2}\laplacian}{\mathbf{q'}}\\
 &= \frac{1}{2}\delta_{\mathbf{q}, \mathbf{q'}}\abs{\mathbf{q}}^2 \\
\braket{\mathbf{q}}{\hat{K}\psi} &= \frac{\hbar^2 q^2}{2m_e} \psi(\mathbf{q})
\end{align}
\subsubsection{Truncation of Reciprocal Space: G-Space}
The kinetic energy operator in reciprocal basis is $\frac{q^2}{2}\dyad{\mathbf{q}}$ and the plane waves $\ket{\mathbf{q}}$ form the operator's eigenbasis.
With the eigenvalues given by $\frac{q^2}{2}$, this operator will dominate the Hamiltonian beyond a certain $q$, translating to lowest energy eigenkets having near negligible components of large $\ket{\mathbf{q}}$ plane-waves.
Therefore, as previously mentioned, an energy cutoff $E_{\textrm{cut}}$ is chosen to truncate the reciprocal space to within the sphere $q \leq \sqrt{2 E_{\textrm{cut}}}$. This truncated reciprocal space will be referred to as \textbf{G-space} from here on as it contains the lattice points of the reciprocal lattice which are commonly referred to as G-vectors. Although the basis is truncated, the actual evaluation of Fourier Transforms would still rely on FFT operations involving a fixed regular grid. But, the components outside of the cutoff will be discarded.

\subsubsection{Implementation: \texttt{GSpaceBase}}
Transformation of quantities such as wavefunctions, charges, potentials, etc. between the two basis is one of the most commonly used operations in plane-wave codes.
Therefore, an easy-to-use and robust mechanism to transform between the two basis is critical for \texttt{Quantum MASALA}.
With multiple FFT libraries available to use in Python, there is also a need for the implementation to be modular.
In \texttt{Quantum MASALA}, the \texttt{GSpaceBase} class fully encapsulates the previously discussed FFT operation within a truncated reciprocal space. Instantiated using a list of G-vectors defining the G-space, it implements the following methods:

\begin{itemize}
\item \texttt{GSpaceBase.allocate\_array()} Creates arrays that are optimal for the FFT routines. Supports library-specific constraints 
\item \texttt{GSpaceBase.check\_array\_type()} Ensures that the arrays are of correct shape and type and is suitable for the selected FFT Library
\item \texttt{GSpaceBase.r2g()} Forward Fourier Transform to truncated G-Space
\item \texttt{GSpaceBase.g2r()} Inverse Fourier Transform from G-Space to Real-Space
\end{itemize}

The \texttt{GSpace} class wraps \texttt{GSpaceBase} with an initialization routine that generates the list of G-vectors from an input kinetic energy cutoff.
\begin{pycode}
import numpy as np
from qtm.containers.field import FieldGType, FieldRType,\
                                 get_FieldG, get_FieldR
from qtm.lattice import ReciLattice
from qtm.gspace.gspc import GSpace
from qtm.constants import RYDBERG

ecutwfn = 40 * RYDBERG
ecutrho = 4 * ecutwfn

# Creating a GSpace instance with KE cutoff E_cut = 160 Ry
recilat = ReciLattice.from_reallat(reallat)
gspc = GSpace(recilat, ecutrho)
print(gspc.grid_shape, gspc.size_g, gspc.size_r)
## (18, 18, 18) 2243 5832

# Printing the G-vectors of gspc
with np.printoptions(
    edgeitems=5, threshold=100, formatter={
    'float': lambda num: f'{num:5.1f}',
    'int': lambda num: f'{float(num):5.1f}',
}):
    for i in range(3):
        print(f'G_{i+1}: ', gspc.g_cryst[i])
    print(f'G^2: ', gspc.g_norm2)
## G_1:  [  0.0   1.0   2.0   3.0   4.0 ...  -5.0  -4.0  -3.0  -2.0  -1.0]
## G_2:  [  0.0   0.0   0.0   0.0   0.0 ...  -1.0  -1.0  -1.0  -1.0  -1.0]
## G_3:  [  0.0   0.0   0.0   0.0   0.0 ...  -1.0  -1.0  -1.0  -1.0  -1.0]
## G^2:  [  0.0   3.0  12.1  27.2  48.4 ...  78.7  51.5  30.3  15.1   6.1]

# Creating a array representing a random value scalar field
FieldR_rho : FieldRType = get_FieldR(gspc)
sfield_r = FieldR_rho.zeros(1)
sfield_r[:] = np.asarray(np.random.rand(gspc.size_r) \
                         + 1j * np.random.rand(gspc.size_r), 
                         like=sfield_r._data)

# Transforming to G-Space using gspc instance
sfield_g = sfield_r.to_g()

# The above is equivalent to 3D forward FFT followed by 
# taking components corresponding to reciprocal
# lattice points that are within the energy cutoff
# print(sfield_r.data.reshape(gspc.grid_shape))
print(np.allclose(
    sfield_g.data, np.fft.fftn(
        sfield_r.data.reshape(gspc.grid_shape)
    ).take(gspc.idxgrid)
))
## True
\end{pycode}

\subsubsection{Optimizing FFT operations for G-Space}
\begin{figure}[!htb]%
    \centering
    \subfloat[\centering G-Space in Cartesian coords]{{\includegraphics[width=7cm]{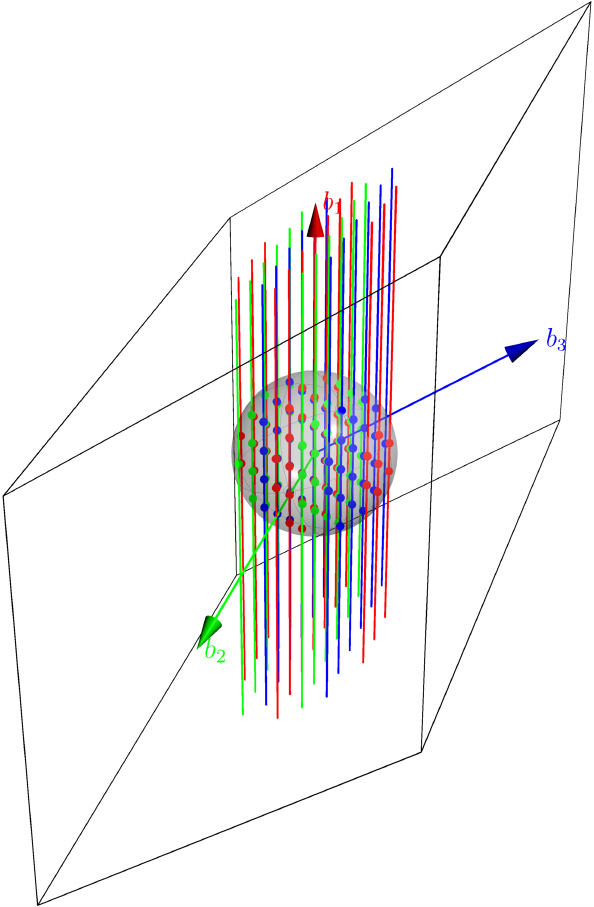} }}%
    \qquad
    \subfloat[\centering G-Space in crystal coords]{{\includegraphics[width=7cm]{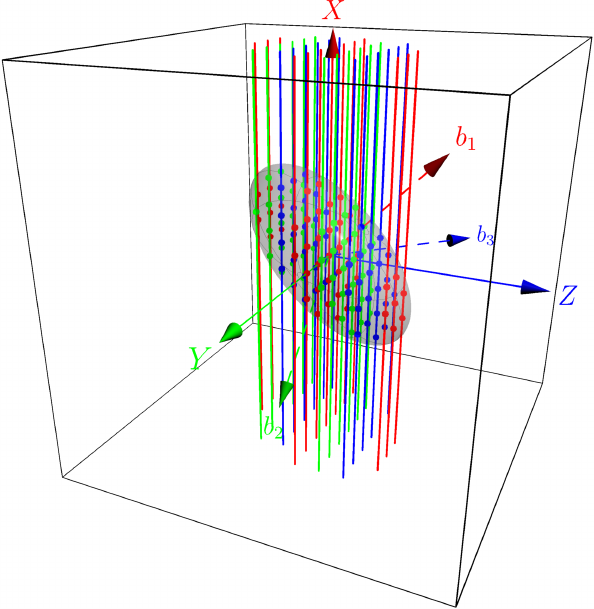} }}%
    \caption{A visual representation of the G-space of a BCC Iron lattice with lattice parameter 5.107 a.u. and energy cutoff 25 Ry.
    FFT along X-Axis is performed only along the sticks that contain atleast one G-vector within cutoff energy}%
    \label{fig1}
\end{figure}

Implementing Fourier Transforms to and from the G-space using a single 3D FFT operation might not be optimal as some Fourier components can be ignored.
By the definition of G-space, certain regions of the FFT grid are set to zero and this region is fixed for a given lattice and cutoff parameters.
When computing the 3D FFT of let's say a $N\times N\times N$ array, we can evaluate them with $N^2$ 1D FFT's across each of the three axes.
Knowing the truncated region, we can skip calling the actual 1D FFT routines on the 1D `sticks' that contain no non-zero elements \cite{AndrewCanning_FFT}. Figure (\ref{fig1}), shows the G-space for the BCC Iron lattice. As can be seen in the figure, a large fraction of the elements are zero.
In theory, this reduces the total number of FFT operations as long as the G-vectors are compactly grouped within the FFT Grid, which is the case when truncating with an energy cutoff. 
But, in practice, the implementation requires data to be rearranged between each 1D FFT calls. Also, 3D FFT implementations involve batched FFT calls where the library operates on multiple 1D arrays simultaneously. These two factors lowers the method's overall efficiency and limits the algorithm effectiveness to only large grid sizes.

The runtime performance is sensitive to its implementation details and highly dependent on the structure of the G-space. But for wave-functions where the number of G-vectors is much smaller than the total number of FFT Grid points, this method reduces the computation involved in \texttt{r2g()}/\texttt{g2r()} methods.

\subsection{Data Containers: \texttt{Buffers}, \texttt{Fields} and \texttt{Wavefun}}
For ease-of-use, we need to pair \texttt{GSpace} with a data container for representing and operating on with quantities like potentials, charge-density etc. that share the same periodicity as the crystal.
A simple NumPy array would not be sufficient as it will not contain any data regarding the lattice and the G-space, which
is required for many methods such as computing the Hartree potential, evaluating the del operator on a scalar/vector field, constructing the kinetic energy operator, etc.
But, the container should mimic a lot of NumPy features like array slicing, indexing, broadcasting, list comprehension, etc. and binary operations like +, -, *, /, \%\, etc.

Therefore, we have implemented the \texttt{Buffer} class that is a custom array-like container that enables the aforementioned functionalities.
Initialized with a \texttt{GSpace} instance and NumPy array, it also provides FFT routines to convert between the two bases.
The \texttt{Buffer} class is extended by the following pairs of classes:
\begin{itemize}
    \item \texttt{FieldR} and \texttt{FieldG}: Represents multidimensional fields
    \item \texttt{WavefunR} and \texttt{WavefunG}: Represents single particle Bloch wavefunctions for a given k-point and stores its periodic part.
\end{itemize}
The \texttt{Buffer} base class implements the following methods:
\begin{itemize}
    \item \texttt{copy()}: creating copies of existing instances
    \item \texttt{empty(gspc, shape)} and \texttt{zeros(gspc, shape)}: Creating an empty/zero array with basis defined by \texttt{gspc} and input shape.
    \item \texttt{to\_r()} / \texttt{to\_g()}: conversion from G-Space to R-Space and vice-versa via respectively.
\end{itemize}
Internally, \texttt{Buffer} supports a select set of NumPy functions that operate on one element of an array at a time. Called as Universal Functions in NumPy, they provide a wide range of routines of which the \texttt{Buffer} class supports the following:
\begin{itemize}
    \item Unary operations like computing trigonometric functions, negation, conjugation, logical/bit-wise inversion, etc.
    \item Binary operations such as +, -, *, /, \% and other algebraic, logical and bit-wise operations.
    \item Reduction operations where one of the axes of the input array is reduced by repeatedly applying a binary operation to the values in array along the given axis.
    \item Matrix multiplication (only across the last two dimensions of array(s))
\end{itemize}

To improve performance, the \texttt{WavefunG} implements the \texttt{WavefunG.vdot(psi, phi)} method for evaluating the dot product $\braket{\psi_i}{\varphi_j}$ that directly calls the ZGEMM BLAS method in \\\texttt{scipy.linalg.blas} module.
This is to prevent calling \texttt{ndarray.conj()} method in NumPy that results in creating a new array and increasing memory usage.

\begin{pycode}
FieldG_gspc : FieldGType = get_FieldG(gspc)

def print_info(f):
    typ = type(f)
    basis = getattr(f, 'basis_type', 'N\A')
    data = getattr(f, 'data', None)
    shape = data.shape if data is not None else 'N/A'
    print(f"type: {str(type(f)):35}\n"
          f"basis: '{basis}', data.shape: {shape}\n")

# Creating a empty GField instance
a_g = FieldG_gspc.empty((4, 10))
print_info(a_g)
## type: <class 'qtm.containers.field.get_FieldG.<locals>.FieldG'>
## basis: 'g', data.shape: (4, 10, 2243)

# Converting to RField instance
a_r = a_g.to_r()
print_info(a_r)
## type: <class 'qtm.containers.field.get_FieldR.<locals>.FieldR'>
## basis: 'r', data.shape: (4, 10, 5832)

# Indexing
arr_idx = a_r[2]
print_info(arr_idx)
## type: <class 'qtm.containers.field.get_FieldR.<locals>.FieldR'>
## basis: 'r', data.shape: (10, 5832)

# Slicing
arr_sl = a_r[:, :5]
print_info(arr_sl)
## type: <class 'qtm.containers.field.get_FieldR.<locals>.FieldR'>
## basis: 'r', data.shape: (4, 5, 5832)

# Unpacking
x, y = a_r[0, :2]
print_info(x)
print_info(y)
## type: <class 'qtm.containers.field.get_FieldR.<locals>.FieldR'>
## basis: 'r', data.shape: (5832,)
## type: <class 'qtm.containers.field.get_FieldR.<locals>.FieldR'>
## basis: 'r', data.shape: (5832,)

# Binary operations
z = x + y
print_info(z)
## type: <class 'qtm.containers.field.get_FieldR.<locals>.FieldR'>
## basis: 'r', data.shape: (5832,)

# Binary operations with broadcasting
z = x + a_r[:2,:3]
print_info(z)
## type: <class 'qtm.containers.field.get_FieldR.<locals>.FieldR'>
## basis: 'r', data.shape: (2, 3, 5832)

# Operations between mismatched basis; raises exception
print(type(x))
z = x + x.to_g()
## TypeError: operand type(s) all returned NotImplemented
## from __array_ufunc__(<ufunc 'add'>, '__call__',  ...
\end{pycode}

\subsection{GPU Acceleration in \texttt{Quantum MASALA}: Writing device-agnostic routines}
Although GPUs have a much higher theoretical compute performance than CPUs, porting CPU code to GPU is not easy, simply due to the difference in their architectures.
For most applications, there is a need to write specialized code that is optimized for running in GPU, resulting in two separate code bases for running in each device.
But, in Python, CuPy (GPU) provides an NumPy-like (CPU) interface to most linear algebra routines, allowing users to quickly port their NumPy codes for GPU Acceleration.
This, paired with the Python array API standard\cite{ArrayAPI}, enables NumPy routines to work with GPU Arrays by implicitly calling the corresponding CuPy routines.
For instance, the keyword argument \texttt{like} in NumPy array creation method can be used to create arrays of the same type as the value passed to the argument, allowing us to write routines using NumPy functions that is also compatible with input CuPy arrays.

With this, \texttt{Quantum MASALA} aims to maximize the reuse of code for GPU routines and one of the key mechanisms implemented to achieve this goal is baked in the \texttt{GSpace} class; handling buffer creation on device and performing FFT operations on device.
This enables one to reuse routines that contain CuPy-aware NumPy functions and perform the same operation(s) in the GPU by just passing the GPU instance. 
For instance, if the use of GPU is enabled in the calculation, the \texttt{RealLattice} constructor object creates a CuPy array-based \texttt{latvec}. Subsequently, all the classes which use that \texttt{RealLattice} object in their constructors will get instantiated on the GPU.

\begin{pycode}
"""cupy_example.py"""
import numpy as np
from qtm import qtmconfig
from qtm.constants import RYDBERG
from qtm.containers.field import get_FieldG
from qtm.crystal.basis_atoms import BasisAtoms
from qtm.crystal.crystal import Crystal
from qtm.gspace.gspc import GSpace
from qtm.kpts import gen_monkhorst_pack_grid
from qtm.lattice import RealLattice
from qtm.pseudo.upf import UPFv2Data


def check_array_types():
    """Function to illustrate the types of arrays used 
    in Quantum MASALA objects"""
    
    # The __init__ method of Lattice class checks if the GPU is enabled 
    # and sets the NDArray to cp.ndarray if it is enabled
    reallat = RealLattice.from_alat(
        alat=10.2,           # Bohr
        a1=[-0.5, 0.0, 0.5], 
        a2=[0.0, 0.5, 0.5], 
        a3=[-0.5, 0.5, 0.0],  
    )

    print(f"{type(reallat.latvec)=}")

    # This information is propagated to all the classes 
    # that are constructed using the reallat object
    si_oncv = UPFv2Data.from_file("Si_ONCV_PBE-1.2.upf")
    si_atoms = BasisAtoms(
        "si",
        si_oncv,
        28.086,
        reallat,
        np.array([[0.875, 0.875, 0.875], [0.125, 0.125, 0.125]]).T,
    )
    print(f"{type(si_atoms.r_cryst)=}")

    crystal = Crystal(reallat, [si_atoms])  # Represents the crystal

    kpts = gen_monkhorst_pack_grid(crystal, (4,4,4), (True,True,True))
    print(f"{type(kpts.k_cryst)=}")
    print(f"{type(kpts.k_weights)=}")
    
    grho = GSpace(crystal.recilat, 100 * RYDBERG)
    print(f"{type(grho.g_cryst)=}")
    
    is_spin = False
    rho_start = get_FieldG(grho).zeros(1 + is_spin)
    print(f"{type(rho_start.data)=}")


# Use set_gpu to enable/disable GPU usage in Quantum MASALA
qtmconfig.set_gpu(True)
print("GPU enabled:", qtmconfig.gpu_enabled)
check_array_types()
# GPU enabled: True
# type(reallat.latvec)=<class 'cupy.ndarray'>
# type(si_atoms.r_cryst)=<class 'cupy.ndarray'>
# type(kpts.k_cryst)=<class 'cupy.ndarray'>
# type(kpts.k_weights)=<class 'cupy.ndarray'>
# type(grho.g_cryst)=<class 'cupy.ndarray'>
# type(rho_start.data)=<class 'cupy.ndarray'>

qtmconfig.set_gpu(False)
print("\nGPU enabled:", qtmconfig.gpu_enabled)
check_array_types()
# GPU enabled: False
# type(reallat.latvec)=<class 'numpy.ndarray'>
# type(si_atoms.r_cryst)=<class 'numpy.ndarray'>
# type(kpts.k_cryst)=<class 'numpy.ndarray'>
# type(kpts.k_weights)=<class 'numpy.ndarray'>
# type(grho.g_cryst)=<class 'numpy.ndarray'>
# type(rho_start.data)=<class 'numpy.ndarray'>

\end{pycode}








\subsection{Parallelization in \texttt{Quantum MASALA}}
\texttt{Quantum MASALA} utilizes Message Passing Interface (MPI) routines provided by the \texttt{mpi4py} Python library to run across multiple processor cores. It can scale across hundreds of processor cores and contains multiple parallelization schemes for effectively distributing its calculation in parallel. Although \texttt{Quantum MASALA}'s implementations are written for parallel operations, it can still be run serially without installing \texttt{mpi4py}.

In order to achieve that, we have wrapped the core objects and the required operations in \texttt{mpi4py}, allowing the library to be an \textbf{optional} run-time dependency. We begin with a brief description about the wrapper and its usage.
\subsubsection{\texttt{QTMComm}: Wrapping MPI Communicators}
`Communicator' is the key object in MPI that connects groups of processes and provides an interface for communicating data across them.
In \texttt{mpi4py}, this is encapsulated by \\\texttt{mpi4py.MPI.Comm} instances that are generated from a given group of processes.
These instances provide methods for communicating data across members of the corresponding group.
In MPI, the group containing all processes is called the `world' and its corresponding communicator is accessed by \texttt{mpi4py.MPI.COMM\_WORLD} in \texttt{mpi4py}.

In order to write parallel code while maintaining the status of the library as an optional dependency, we have implemented a wrapper class named \texttt{QTMComm} that has a similar interface to \texttt{mpi4py.MPI.Comm}.
It only imports \texttt{mpi4py} if available and if not available, it assumes serial operation and the wrapper will not utilize MPI routines at all while providing consistent behaviour.
We have also extended the class to provide a few group operations such as \texttt{MPI\_Group\_split} and \texttt{MPI\_Group\_incl}.

We have also implemented context managers that serve the following purpose:
\begin{itemize}
\item It provides a visual cue that indicates parallelism i.e code within a \texttt{with} block show that the operation is across multiple processes.
\item At the end of the \texttt{with} block, the Barrier operation \texttt{MPI\_Comm\_barrier} is called automatically, synchronizing all processes in the communicator
\item The group operations are implemented with an additional boolean argument \\\texttt{sync\_with\_parent} that specifies if the barrier operation is also applied to the parent group when exiting the \texttt{with} block.
\end{itemize}
The last feature is especially handy in situations where you need only a subgroup of processes to execute a code-block while the remainder waits for their completion. Below, we will illustrate some use cases of \texttt{QTMComm}'s context manager.

\subsubsection{Example 1: Using \texttt{QTMComm}'s Context Manager}

\begin{pycode}
"""comm_example.py"""
import time

from mpi4py.MPI import COMM_WORLD
from qtm.mpi import QTMComm

# Creating 'QTMComm' instance from 'COMM_WORLD'
comm_world = QTMComm(COMM_WORLD)
wrld_size, wrld_rank = comm_world.size, comm_world.rank

# Function to add timestamp before printing to stdout
def print_msg(msg: str):
    curr_time = time.strftime("%H:%M:%S", time.localtime())
    print(f"{curr_time}: {msg}", flush=True)
    
with comm_world as comm:
    size, rank = comm.size, comm.rank
    print_msg(f"Hello from process #{rank}/{size}")
    # Proc 1 going to sleep while the rest skip ahead
    if comm.rank == 1:
        print_msg(f"process #{rank} going to sleep for 3 seconds")
        time.sleep(3)
    print_msg(f"process #{rank} is at the end of 'with' code-block")

# When exiting, all procs in comm will be in sync.
# No need for 'comm.Barrier()' here.
# So all procs will print the message below at the exact time.
print_msg(f"process #{rank} has exited 'with' code-block")
\end{pycode}
\paragraph{Output}
\begin{verbatim}
$ mpirun -n 4 python comm_example.py
06:04:28: Hello from process #0/4
06:04:28: process #0 is at the end of 'with' code-block
06:04:28: Hello from process #1/4
06:04:28: process #1 going to sleep for 3 seconds
06:04:28: Hello from process #2/4
06:04:28: process #2 is at the end of 'with' code-block
06:04:28: Hello from process #3/4
06:04:28: process #3 is at the end of 'with' code-block
06:04:31: process #1 is at the end of 'with' code-block
06:04:31: process #1 has exited 'with' code-block
06:04:31: process #0 has exited 'with' code-block
06:04:31: process #2 has exited 'with' code-block
06:04:31: process #3 has exited 'with' code-block
\end{verbatim}

\subsubsection{Example 2: Dividing Communicator using \texttt{Split}}
\begin{pycode}
# Processes are divided into two groups
# rank    0  1  2  3
colors = [0, 1, 1, 0]  # Group number
keys   = [1, 1, 0, 0]  # Rank in group

wrld_size = comm_world.size
wrld_rank = comm_world.rank

c = colors[wrld_rank]
k = keys[wrld_rank]

with comm_world.Split(c, k) as comm:
    grp_id = c
    grp_size, grp_rank = comm.size, comm.rank
    print(f"process #{wrld_rank}/{wrld_size} is assigned to "
          f"subgroup #{c} and its rank is "
          f"{grp_rank}/{grp_size}")
\end{pycode}
\paragraph{Output}
\begin{verbatim}
process #0/4 is assigned to subgroup #0 and its rank is 1/2
process #1/4 is assigned to subgroup #1 and its rank is 1/2
process #2/4 is assigned to subgroup #1 and its rank is 0/2
process #3/4 is assigned to subgroup #0 and its rank is 0/2
\end{verbatim}

\subsubsection{Example 3: Running a code-block on a subset of processes using \texttt{Init}}
\begin{pycode}
# Selecting the processess to include in subgroup
grp_iproc = [0, 3]

with comm_world.Incl(grp_iproc) as comm:
    if not comm.is_null:
        # The newly created group will be active
        grp_size, grp_rank = comm.size, comm.rank
        print_msg(f"process #{wrld_rank}/{wrld_size} is part of the "
                  f"subgroup and its rank is {grp_rank}/{grp_size}")
        print_msg(f"process #{wrld_rank} going to sleep for 3 seconds")
        time.sleep(3)
    else:
        # while the remaining processes wait at the end of the
        # 'with' codeblock
        print_msg(f"process #{wrld_rank}/{wrld_size} is not part of "
                  "the subgroup")
    print_msg(f"process #{wrld_rank} is at the end of 'with' code-block")

# When exiting, all procs in comm will be in sync.
# No need for 'comm.Barrier()' here.
# So all procs will print the message below at the exact time.
print_msg(f"process #{comm_world.rank} has exited 'with' code-block")
\end{pycode}
\paragraph{Output}
\begin{verbatim}
06:19:32: process #1/4 is not part of the subgroup
06:19:32: process #1 is at the end of 'with' code-block
06:19:32: process #2/4 is not part of the subgroup
06:19:32: process #2 is at the end of 'with' code-block
06:19:32: process #0/4 is part of the subgroup and its rank is 0/2
06:19:32: process #0 going to sleep for 3 seconds
06:19:32: process #3/4 is part of the subgroup and its rank is 1/2
06:19:32: process #3 going to sleep for 3 seconds
06:19:35: process #0 is at the end of 'with' code-block
06:19:35: process #3 is at the end of 'with' code-block
06:19:35: process #0 has exited 'with' code-block
06:19:35: process #3 has exited 'with' code-block
06:19:35: process #2 has exited 'with' code-block
06:19:35: process #1 has exited 'with' code-block
\end{verbatim}

\subsubsection{Implemented Parallelization Modes}
Using \texttt{QTMComm}, \texttt{Quantum MASALA} implements different levels of parallelization, which can be labelled as follows:
\begin{itemize}
    \item Parallelism over \textbf{k-points}: Quantities that depend on only one or two k-points can be evaluated in parallel by distributing the k-points across subgroups called k-groups.
    Each k-group is assigned a subset of k-points/k-point pairs and are allowed to run independently till they are done with their assigned tasks.
    Communication across different k-groups is required only when the quantity is summed/reduced across k-points.
    \item Parallelism over \textbf{bands}: Subroutines that operate on a set of bands such as solving the Kohn-Sham Hamiltonian, computing the electron density from Bloch states, etc. usually involves operations that are applied on a batch of wave-functions which can be done in parallel like, for instance, evaluating $\operatorname{\hat{H}_{\textrm{KS}}}\ket{\psi}$.
    The k-group is further split into band-groups, which are assigned a subset of wave-functions to operate one.
    It must be noted that this parallelism is communication and usually memory intensive.
    \item Parallelism over \textbf{PW Basis}: This parallelism involves the distribution of the plane-wave basis/G-Space and its Real-Space dual. Implementation requires distributed FFT routines and frequent transfer of data between processes, making it the most communication-heavy scheme. But, unlike the band-distributed parallelization, the memory is distributed across processes, making it effective for large-size crystals and supercell calculations. Currently, this parallelization level cannot be used simultaneously with the other two, but it will be made compatible with other parallelization levels in future updates to the code.
\end{itemize}

\section{Code layout}
\label{sec:org960962c}
Apart from the core structures discussed in the previous section, the implementation of electronic-structure methods in \texttt{Quantum MASALA} is split into different sub-directories based on its scope, each forming its own submodules.
In the following section, each module and its contents will be described in brief.

\subsection{`pseudo' module}
\label{sec:org84c96f5}
The \texttt{qtm.pseudo} module deals with constructing the (pseudo)-potential in the crystal from the given pseudopotential data. This potential represents the interaction between the valence electrons and the ions (nucleus + core electrons) in the crystal. \texttt{Quantum MASALA} implements the Norm-Conserving PseudoPotential (NCPP) and supports Optimized Norm-Conserving Vanderbilt (ONCV) Pseudpotentials\cite{ONCV} which are relatively simpler to implement compared to projector-augmented waves (PAWs)\cite{PAW} and ultrasoft pseudopotentials\cite{USPP}, while yielding accurate results\cite{NCPPAcc, PseudoDojo, DeltaBench}

The pseudopotentials are evaluated in two parts:
\begin{itemize}
\item the local term generated using function \texttt{loc\_generate\_pot\_rhocore} which also gives density of core electrons for non-local core correction (NLCC) in computing XC potentials.
\item the non-local term generated using instances of class \texttt{NonlocGenerator}
\end{itemize}

This module also contains the function \texttt{loc\_generate\_rhoatomic} which is used by \texttt{qtm.core.rho\_atomic} to construct electron densities from superposition of atomic charges.

Currently, \texttt{Quantum MASALA} supports Unified Pseudopotential Format (UPF) v2 files as input, which is processed when specifying the basis atoms of crystal using \texttt{qtm.core.BasisAtoms}.
The SG15\cite{SG15} ONCV set of psuedopotential is recommended as it has been extensively tested with \texttt{Quantum MASALA} as part of the DFT benchmarks.

\subsection{`dft' module}
\label{sec:org73e51fc}
The \texttt{dft} module of \texttt{Quantum MASALA} implements the solvers of the KS Hamiltonian and the SCF iteration routines.

For computing the ground-state of the KS system, only the lowest occupied states of the Hamiltonians are required, which constitutes a small fraction of the eigenvalues of the Hamiltonian matrix (\ref{KSHamMat}).
\texttt{Quantum MASALA} currently implements the Davidson Method with Overlap, which is based on the same from \texttt{Quantum ESPRESSO}.
A key feature of these methods is that they do not require the explicit construction of the matrix that is being diagonalized.
The routines instead involve repeated application of the matrix on `guess' vectors.
The algorithms treat the linear transformation of input vectors as a black box that is consistent with the requirements of a Hermitian operator.
\texttt{Quantum MASALA} implements the Hamiltonian as a \texttt{class KSHam} instance that contains the \texttt{h\_psi()} method which returns the \(\hat{H}_{\textrm{KS}}(\mathbf{k})\ket{\psi_{\mathbf{k}}}\) for an input trial vector \(\ket{\psi_{\mathbf{k}}}\).

The implemented Davidson Method can operate on either NumPy (CPU) and CuPy (GPU) arrays. This is possible due to the close interoperability between the two libraries.

The SCF iteration requires mixing of charge densities to accelerate convergence. \\\texttt{Quantum MASALA} contains multiple mixing methods, the default being Broyden mixing (Mixed\cite{ModBroyden} and General\cite{GenBroyden}). As these algorithms uses the charge densities computed from previous iterations, these routines are implemented as classes with its own memory and a method for generating the charge density for the next iteration.

When running across multiple processes via \texttt{mpiexec}, \texttt{Quantum MASALA} implements two levels of parallelism.
The first involves the distribution of input $\mathbf{k}$-points across different process groups (k-group). This is effective when working with large sets of $\mathbf{k}$-points and scales linearly.
The second distributes the \texttt{h\_psi()} calls across processes in the k-group, effectively parallelizing across bands. This is useful when dealing with supercells which result a large number of occupied states.
Parallelism through OpenMP is not controlled by \texttt{Quantum MASALA} and instead relies on environmental variables that specify number of threads the linked BLAS and FFT Libraries can use.

\begin{pycode}
import numpy as np
from qtm.config import qtmconfig
from qtm.constants import ELECTRONVOLT, RYDBERG
from qtm.crystal import BasisAtoms, Crystal
from qtm.dft import DFTCommMod, scf
from qtm.gspace import GSpace
from qtm.io_utils.dft_printers import print_scf_status
from qtm.kpts import gen_monkhorst_pack_grid
from qtm.lattice import RealLattice
from qtm.logger import qtmlogger
from qtm.mpi import QTMComm
from qtm.pseudo import UPFv2Data
from mpi4py.MPI import COMM_WORLD


world_comm = qtmconfig.get_world_comm()

# Only k-pt parallelization:
dftcomm = DFTCommMod(comm_world, comm_world.size, 1)
# Only band parallelization:
# dftcomm = DFTCommMod(comm_world, 1, 1)

# Lattice
reallat = RealLattice.from_alat(alat=5.1070,  # Bohr
                                a1=[ 0.5,  0.5,  0.5],
                                a2=[-0.5,  0.5,  0.5],
                                a3=[-0.5, -0.5,  0.5])

# Atom Basis
fe_oncv = UPFv2Data.from_file('Fe_ONCV_PBE-1.2.upf')
fe_atoms = BasisAtoms('fe', fe_oncv, 55.487, reallat, np.array(
    [[0., 0., 0.]]
).T)

crystal = Crystal(reallat, [fe_atoms, ])  # Represents the crystal
print(crystal)


# Generating k-points from a Monkhorst Pack grid 
# (reduced to the crystal's IBZ)
mpgrid_shape = (8, 8, 8)
mpgrid_shift = (True, True, True)
kpts = gen_monkhorst_pack_grid(crystal, mpgrid_shape, mpgrid_shift)
print(kpts)

# -----Setting up G-Space of calculation-----
ecut_wfn = 40 * RYDBERG
ecut_rho = 4 * ecut_wfn
grho = GSpace(crystal.recilat, ecut_rho)
gwfn = grho


# -----Spin-polarized (collinear) calculation-----
is_spin, is_noncolin = True, False
# Starting with asymmetric spin distribution else convergence
# may yield only non-magnetized states
mag_start = [0.1]
numbnd = 12  # Ensure adequate # of bands if system is not an insulator

# Control parameters that specify how occupation of eigenstates
# are computed
occ = 'smear'
smear_typ = 'gauss'
e_temp = 1E-2 * RYDBERG

conv_thr = 1E-8 * RYDBERG
diago_thr_init = 1E-2 * RYDBERG

# Calling the scf routine
out = scf(dftcomm, crystal, kpts, grho, gwfn,
          numbnd, is_spin, is_noncolin,
          rho_start=mag_start, 
          occ_typ='smear', smear_typ='gauss', e_temp=e_temp,
          conv_thr=conv_thr, diago_thr_init=diago_thr_init,
          iter_printer=print_scf_status)

# Unpacking the output of scf function
scf_converged, rho, l_wfn_kgrp, en = out

if comm_world.rank == 0:
    print("SCF Routine has exited")
    print(qtmlogger)
\end{pycode}

\subsection{`tddft' module}
The \texttt{qtm.tddft\_gamma} module implements propagation of the TDKS equations, which enables solving for the
optical spectrum of the system. The module is designed for molecular systems, where the molecule in question is placed
in a box large enough to isolate it from other molecules in neighbouring unit cells in the crystal. The calculations are
limited to the gamma point $\Gamma$ of the reciprocal lattice.

Propagating the TDKS equation involves:
\begin{enumerate}
\item Constructing an accurate approximation to $\hat{U}(t+\Delta t, t)$, which usually involves implementing the exponentiation of the Hamiltonian
\item An accurate scheme to solve the TDKS equation and propagate the states from $t$ to $t + \Delta t$
\end{enumerate}
In \texttt{Quantum MASALA}, the TDDFT module is split into 3 major sections:
\begin{itemize}
\item \texttt{expoper} submodule: implements the exponentiation of the Hamiltonian $\exp{-i\Delta t \hat{H}}$ by extending the \texttt{KSHam} class in DFT module. Currently, two implementations are available:
\begin{itemize}
\item Taylor Series Expansion: truncated to order $k$.
\begin{equation}
\exp{-i\Delta t \hat{H}} \approx \sum_{r=0}^{k} \frac{(-i\Delta t)^r}{r\,!}\hat{H}^{r}
\end{equation}

\item Split-Operator
\begin{equation}
\exp{-i\Delta t \hat{H}} \approx \exp{\frac{-i\Delta t}{2}\hat{T}}\exp{-i\Delta t \hat{V}}\exp{\frac{-i\Delta t}{2}\hat{T}}
\end{equation}
\end{itemize}

Like \texttt{KSHam}, the operators here are not explicitly constructed as a dense matrix, but implemented as a routine, named \texttt{prop\_psi()}, that returns the transformed vectors of input wavefunctions.

\item \texttt{prop} submodule: contains implementations that use the propagators defined in \texttt{expoper} submodule and solves the TDKS equation for a given time step $t\rightarrow t+\Delta t$. Currently, this submodule contains
\begin{itemize}
\item Enforced Time-Reversal Symmetry (ETRS) Propagator: 
\begin{align}
\bar{\psi}_{n+1}&=\exp{-i\delta t \hat{H}[\psi_n]}\psi_n \\
\psi_{n+1} &= \exp{-i\delta t \hat{H}[\bar{\psi}_{n+1}] / 2}\exp{-i\delta t \hat{H}[\psi_n]/2}\psi_n
\end{align}

\item Split-Operator Propagator: Application of the split-operator exponential. Not time-reversal symmetric.

\end{itemize}
\item \texttt{propagate} routine: combines the above two parts and solves the TDKS equations for the given time period. The propagation method used is specified by \\\texttt{config.tddft\_exp\_method} and \texttt{config.tddft\_prop\_method}. At each time step, a user-specified callback function is executed where quantities like dipole moment can be computed from the time-varying charge density and wavefunctions.
\end{itemize}

Using the above, the calculation of optical absorption spectra is implemented in \texttt{optical.py}. Parallelization across bands is supported in this module. Using libraries compiled with multithreading enabled can speed up calculation. GPU Acceleration is currently not implemented as the Python wrapper to Libxc does not support CuPy arrays.

\begin{pycode}
# TDDFT calculation requires the system's
# ground state charge and wavefun
# So the SCF routine is performed first
# followed by optical response calculation.

# ---------- SCF Calculation -----------
import os
import numpy as np
from mpi4py.MPI import COMM_WORLD
from qtm.config import qtmconfig
from qtm.constants import ELECTRONVOLT, RYDBERG
from qtm.containers.wavefun import get_WavefunG
from qtm.crystal import BasisAtoms, Crystal
from qtm.dft import DFTCommMod, scf
from qtm.gspace import GSpace
from qtm.io_utils.dft_printers import print_scf_status
from qtm.kpts import KList
from qtm.lattice import RealLattice
from qtm.logger import qtmlogger
from qtm.mpi import QTMComm
from qtm.pseudo import UPFv2Data
from qtm.tddft_gamma.optical import dipole_response

comm_world = QTMComm(COMM_WORLD)
dftcomm = DFTCommMod(comm_world, 1, comm_world.size)

# Lattice
reallat = RealLattice.from_alat(alat=30.0, # Bohr
                                a1=[1., 0., 0.],
                                a2=[0., 1., 0.],
                                a3=[0., 0., 1.])


# Atom Basis
c_oncv = UPFv2Data.from_file('C_ONCV_PBE-1.2.upf')
h_oncv = UPFv2Data.from_file('H_ONCV_PBE-1.2.upf')

# C atom at the center of the cell
c_atoms = BasisAtoms.from_angstrom('C', c_oncv, 12.011, reallat,
                                  0.529177*np.array([15., 15., 15.]))
coords_ang = 0.642814093
h_atoms = coords_ang * np.array(
    [[ 1,  1,  1],
     [-1, -1,  1],
     [ 1, -1, -1],
     [-1,  1, -1]])
# Shift the H atoms to the center of the cell
h_atoms += 0.529177 * 15.0 * np.ones_like(h_atoms)
h_atoms = BasisAtoms.from_angstrom('H', h_oncv, 1.000, reallat,
                                  *h_atoms)

crystal = Crystal(reallat, [c_atoms, h_atoms])
kpts = KList.gamma(crystal.recilat)
print(kpts.numkpts)


# -----Setting up G-Space of calculation-----
ecut_wfn = 25 * RYDBERG
ecut_rho = 4 * ecut_wfn
grho = GSpace(crystal.recilat, ecut_rho)
gwfn = grho


is_spin, is_noncolin = False, False
numbnd = crystal.numel // 2
occ = 'fixed'
conv_thr = 1E-10 * RYDBERG
diago_thr_init = 1E-5 * RYDBERG


out = scf(dftcomm, crystal, kpts, grho, gwfn,
        numbnd, is_spin, is_noncolin,
        occ_typ=occ,
        conv_thr=conv_thr, diago_thr_init=diago_thr_init,
        iter_printer=print_scf_status)

scf_converged, rho, l_wfn_kgrp, en = out

WavefunG = get_WavefunG(l_wfn_kgrp[0][0].gkspc, 1)

print("SCF Routine has exited")
print(qtmlogger)


# ---------- Determining Optical Response ----------
# Electric field kick (in z-direction) in Hartree atomic units
gamma_efield_kick = 1e-4 
# Time step in Hartree atomic units
time_step = 0.1    
        
numsteps = 10_000

# Choosing methods for propagation and for evaluating matrix exponentials
qtmconfig.tddft_prop_method = 'etrs'
qtmconfig.tddft_exp_method = 'taylor'

# Computing the real-time dipole response
# to a delta perturbation at t=0 along z-Axis
dip_z = dipole_response(comm_world, crystal, l_wfn_kgrp,
                        time_step, numsteps, gamma_efield_kick, 'z')

# Transforming the response to energy spectrum
# NOTE: While the real-time propagation runs,
# One can use the dipz_temp.npy file to 
# plot the dipole response.

en_start = 0
en_end = 40 * ELECTRONVOLT
en_step = en_end / len(dip_z)
damp_func = "gauss"
dip_en = dipole_spectrum(
    dip_t=dip_z,
    time_step=time_step,
    damp_func=damp_func,
    en_end=en_end,
    en_step=en_step,
)
plt.plot(dip_en[0] / ELECTRONVOLT, np.imag(dip_en[1])[:, 2])
plt.savefig("dipz.png")

fname = "dipz.npy"
if os.path.exists(fname) and os.path.isfile(fname):
    os.remove(fname)
np.save(fname, dip_z)
\end{pycode}

\subsection{`gw' module}
The implementation of GW Approximation in the package is based on \texttt{BerkeleyGW} \cite{BGW2012}.
Static-COHSEX and Hybertsen-Louie Plasmon Pole methods of self-energy calculation are currently available in the \texttt{gw} module. As of now, the module supports gapped systems.
Further, \textit{Static-Remainder} method described in \cite{StaticRemainder2013} has been implemented to reduce number of unoccupied bands required for the convergence of Plasmon-Pole self energy values.
The three primary classes in this module are: \texttt{Epsilon}, \texttt{Sigma}, and \texttt{Vcoul}. The classes support \texttt{BerkeleyGW}'s input files, namely \texttt{WFN[q].h5}, \texttt{RHO}, \texttt{vxc[0].dat}, \texttt{epsilon.inp} and \texttt{sigma.inp}.

Input parameters for GW calculations are handled by \texttt{EpsInp} and \texttt{SigmaInp} classes.
The input data can be provided either manually, by constructing the \texttt{EpsInp} object, or by reading \texttt{BerkeleyGW}-compatible input file \texttt{epsilon.inp}.

\begin{pycode}
from qtm.gw.io_bgw.epsinp import Epsinp

# Constructing input object manually
epsinp = Epsinp(epsilon_cutoff=10.0,
                use_wfn_hdf5=True,
                number_bands=8,
                write_vcoul=True,
                qpts=[[0.0, 0.0, 0.001],
                      [0.0, 0.0, 0.2]],
                is_q0=[True, False])
                
# Alternatively, one can read from epsilon.inp file
epsinp = Epsinp.from_epsilon_inp(filename=dirname+'epsilon.inp')

# There is a similar system for reading SigmaInp
from quantum_masala.gw.io_bgw.sigmainp import Sigmainp
sigmainp = Sigmainp.from_sigma_inp(filename=dirname+'sigma.inp')
\end{pycode}

Wavefunction data generated by mean-field codes can be read using the \texttt{wfn2py} utility, which assumes that the incoming data satisfies \texttt{BerkeleyGW}'s \texttt{wfn.h5} specification. The data is stored as a \texttt{NamedTuple} object. Alternatively, the data generated from scf runs within \texttt{Quantum MASALA} can also be passed directly to Epsilon and Sigma constructors in the form of \texttt{qtm} objects as shown later.

We also require wavefunctions on a shifted grid to calculate dielectric matrix at $q\to 0$. This shifted k-grid dataset will be referred to as \texttt{wfnqdata}.

\begin{pycode}
from quantum_masala.gw.io_bgw.wfn2py import wfn2py
wfndata = wfn2py(dirname+'WFN.h5')
wfnqdata = wfn2py(dirname+'WFNq.h5')
\end{pycode}

\subsubsection{Class Epsilon}
Calculates RPA irreducible polarizability matrix $P$, and inverse dielectric matrix $\epsilon^{-1}$ using DFT energy eigenfunctions.

\texttt{Epsilon} class can be initialized by either directly passing the required \texttt{qtm} objects or by passing the input objects discussed earlier.

\begin{pycode}
from qtm.gw.core import QPoints
from qtm.gw.epsilon import Epsilon
from qtm.klist import KList

kpts_gw =   KList(  recilat=kpts.recilat,   
                    cryst=kpts.k_cryst.T,   
                    weights=kpts.k_weights)
                    
kpts_gw_q = KList(  recilat=kpts_q.recilat, 
                    cryst=kpts_q.k_cryst.T, 
                    weights=kpts_q.k_weights)

# Manual initialization
epsilon = Epsilon(
    crystal = crystal,
    gspace = grho,
    kpts = kpts_gw,
    kptsq = kpts_gw_q,        
    l_wfn = l_wfn_kgrp,
    l_wfnq = l_wfn_kgrp_q,
    qpts = QPoints.from_cryst(  kpts.recilat, 
                                epsinp.is_q0, 
                                *epsinp.qpts),
    epsinp = epsinp,
)

# Alternative: Initializing using data objects
epsilon = Epsilon.from_data(wfndata=wfndata, 
                            wfnqdata=wfnqdata, 
                            epsinp=epsinp)
\end{pycode}

The three main steps involved in the calculation have been mapped to corresponding functions:
\begin{enumerate}
    \item \texttt{matrix\_elements}: Calculation of plane wave matrix elements
    \begin{equation}
        M_{nn'}({\textbf k},{\textbf q},{\textbf G}) = \bra{n\,{\textbf k}{+}{\textbf q}}e^{i({\textbf q}+{\textbf G})\cdot{\textbf r}}\ket{n'\,\textbf k}
    \end{equation}
    where the $\textbf G$-vectors included in the calculation satisfy $|\textbf q + \textbf G|^2 < E_{\text{cut}}$. 
    Since this is a convolution in k-space, the time complexity can be reduced from $\mathcal{O}\left(N^2_G\right)$ to $\mathcal{O}\left(N_G\ln N_G\right)$ by using Fast Fourier Transform, where $N_G$  the number of reciprocal lattice vectors in the wavefunction.
    \begin{equation}
    M_{nn'}({\bf k},{\bf q},\{{\bf G}\}) = {\textrm{FFT}}^{-1}\left( \phi^{*}_{n,{\bf k}+{\bf q} }({\bf r}) \phi_{n',{\bf k} }({\bf r}) \right).
    \end{equation}
    where $\phi_{n',{\bf k}}({\bf r}) = {\textrm{FFT}}\left( \psi_{n\bf k}(\bf G)\right)$. 
    \item \texttt{polarizability}: Calculation of RPA polarizability matrix $P$
    \begin{equation}
        P_{\textbf{GG'}}{\left({\textbf q}\;\!;0\right)}=
        \,\,{}\sum_{n}^{\textrm{occ}}\sum_{n'}^{\textrm{emp}}\sum_{{\textbf k}}
        \frac{
        \bra{n'\textbf k}e^{-i({\textbf q}+{\textbf G})\cdot{\textbf r}}\ket{n{\textbf k}{+}{\textbf q}}
        \bra{n{\textbf k}{+}{\textbf q}}e^{i({\textbf q}+{\textbf G'})\cdot{\textbf r}}\ket{n'\textbf k}
        }{E_{n{\textbf k}{+}{\textbf q}}\,{-}\,E_{n'{\textbf k}}}.
    \end{equation}
    \item \texttt{epsilon\_inverse}: Calculation of (static) epsilon-inverse matrix
    \begin{equation}
        \epsilon_{\textbf{GG'}}{\left({\textbf q}\;\!\right)}=
        \delta_{\textbf{GG'}}\,{-}\,v{\left({\textbf q}{+}{\textbf G}\right)} \,
        P_{\textbf{GG'}}{\left({\textbf q}\;\!\right)}
    \end{equation}
    where $ v(\textbf{q} + \textbf{G}) = \frac{8\pi}{\left|\textbf q + \textbf G\right|^2} $ is bare Coulomb potential, written in Rydberg units. If this formula is used as-is for the case where $|\textbf q| = |\textbf G| = 0$, the resulting $v\left({\textbf{q=0}, \textbf{G=0}}\;\!\right)$ blows up as $1/q^2$. However, for 3D gapped systems, the matrix elements $\big| M_{nn'}\left({\bf k},{\textbf{q}\to\textbf{0}},{\textbf{G=0}}\right)\big| \sim q$ cancel the Coulomb divergence and $\epsilon_{\textbf{00}}\left({\textbf q\to\textbf{0}}\;\!\right) \sim q^2/q^2$ which is a finite number. In order to calculate $\epsilon_{\textbf{00}}\left({\textbf q=\textbf{0}}\;\!\right)$, we use the scheme specified in \cite{BGW2012}, wherein $q=0$ is replaced with a small but non-zero value. Since matrix element calculation involves the eigenvectors $\ket{n{\textbf k}{+}{\textbf q}}$, having a non-$\Gamma$-centered $\textbf q\to 0$ point requires mean-field eigenvectors over a shifted $k$-grid.
\end{enumerate}

\begin{pycode}
# Calculate plane wave matrix elements 
mtxel = next(epsilon.matrix_elements(i_q=i_q))

# Calculate polarizability matrix
chimat = epsilon.polarizability(mtxel)

# Calculate epsilon inverse matrix
epsinv = epsilon.epsilon_inverse(i_q = i_q, 
                                 polarizability_matrix = chimat,
                                 store = True)
\end{pycode}


\subsubsection{Class Sigma}
The \texttt{Sigma} class provides methods for calculation of GW self-energy. The self-energy approximations available in the module are: Hartree-Fock bare exchange, Static-COHSEX\cite{Hedin65} and Hybertsen-Louie Plasmon Pole method\cite{HybertsenLouie86}\cite{ZhangLouie89}. 

\begin{pycode}
from qtm.gw.sigma import Sigma

sigma = Sigma.from_qtm_scf(
    crystal = crystal,
    gspace = grho,
    kpts = kpts_gw,
    kptsq=kpts_gw_q,
    l_wfn_kgrp=l_wfn_kgrp,
    l_wfn_kgrp_q=l_wfn_kgrp_q,
    sigmainp=sigmainp,
    epsinp = epsinp,
    epsilon=epsilon,
    rho=rho,
    vxc=vxc
)

# Alternatively, the Sigma object can also be 
# initialized from pw2bgw.x output data 
# (after being processed by wfn2hdf5.x).
sigma = Sigma.from_data(
    wfndata=wfndata, wfnqdata=wfnqdata,
    sigmainp=sigmainp,
    epsinp=epsinp,
    l_epsmats=epsilon.l_epsinv,
    rho=rho,
    vxc=vxc,
    outdir=outdir,
)
\end{pycode}

Conventionally, self-energies are divided into a frequency independent Fock exchange energy term $\Sigma_{\textrm{X}}$, and a correlation term $\Sigma_{\textrm{Corr}}$, which is further divided into a screened exchange (minus bare exchange) part $\Sigma_{\textrm{SX}} \equiv \Sigma_{\textrm{SEX}} - \Sigma_{\textrm{X}}$ and the Coulomb-hole part $\Sigma_{\textrm{COH}}$:
\begin{equation*}
    \Re\left(\Sigma\right) = \Sigma_{\textrm{X}} + \Sigma_{\textrm{SX}} + \Sigma_{\textrm{COH}}
\end{equation*}

Following are the core methods provided by \texttt{Sigma} class: 
\begin{itemize}
    \item Hartree-Fock bare exchange : \texttt{sigma\_x}
    \item Static COHSEX
    \begin{itemize}
        \item Screened exchange : \texttt{sigma\_sx\_static}
        \item Coulomb hole partial sum : \texttt{sigma\_ch\_static}
        \item Coulomb hole exact : \texttt{sigma\_ch\_static\_exact}
    \end{itemize}
    \item Hybertsen Louie Plasmon Pole
    \begin{itemize}
        \item Screened exchange : \texttt{sigma\_sx\_gpp}
        \item Coulomb hole partial sum : \texttt{sigma\_ch\_gpp}
    \end{itemize}
    \item A \texttt{matrix\_elements} function, analogous to that in \texttt{Epsilon} class.
\end{itemize}

\begin{pycode}
# Do a GPP quasiparticle energy calculation
sigma.calculate_gpp()

# Do a static COHSEX quasiparticle energy calculation
sigma.calculate_static_cohsex()

# Individual function calls
sigma.sigma_sx_static()
\end{pycode}


\subsubsection{Class Vcoul}
The \texttt{Vcoul} class provides methods to calculate bare Coulomb interaction, $v(\textbf q, \textbf G) = 8\pi/|\textbf q + \textbf G|^2$, along with methods required to average the potential over mini-Brillouin zones. 

Coulomb potential $\sim 1/|\textbf q + \textbf G|^2$ diverges when $q=G=0$. However, the Coulomb potential values over a discretized $G$-space grid only represent the average potential in the mini-Brillouin zones corresponding to those grid points. As a result we construct, 
$$V_{\textrm{grid}}(\textbf q, \textbf G) = \int_{\textrm{minibz}} \frac{2}{|\textbf q + \textbf G|^2} \text{d}^3\textbf G$$
(in Rydberg units) which turns out to be convergent for all grid points.
Even though the apparent divergence occurs only for the $q=G=0$ case, the averaging is typically performed for all q-points and all G-points for better accuracy\cite{Reproducibility2020}. 

Here we list the primary methods in \texttt{Vcoul} class:
\begin{itemize}
    \item \texttt{v\_bare}: Bare Coulomb potential for a given q-point. 
    \item \texttt{v\_minibz\_montecarlo}: Averaged bare Coulomb potential, where naïve Monte Carlo averaging has been used.
    \item \texttt{v\_minibz\_sphere}: Calculates exact integral of Coulomb potential over a spherical region with the same volume as the corresponding mini-Brillouin zone.
    \item \texttt{v\_minibz\_montecarlo\_hybrid}: Calculates exact integral within a sphere inscribed in the mini-Brillouin zone and uses Monte Carlo averaging for the rest.
\end{itemize}

\begin{pycode}
# Initialization
vcoul = Vcoul( gspace = wfndata.grho, 
               qpts = qpts, 
               bare_coulomb_cutoff = epsinv.epsilon_cutoff)

# Populate with coulomb potential averaged within mini-Brillouin zones
vcoul.calculate_vcoul(averaging_func=vcoul.v_minibz_montecarlo_hybrid)

# Print vcoul in BerkeleyGW format
vcoul.write_vcoul()

# Load vcoul from BerkeleyGW format
vcoul.load_vcoul(filename="vcoul.dat")
\end{pycode}

\section{Results}
In this section, we demonstrate the current capabilities of \texttt{Quantum MASALA} by showing some results obtained from it. Here, we present benchmarking results that show the accuracy of the implemented calculations in the code. We also compare its consistency and performances with standard packages like \texttt{Quantum ESPRESSO} for DFT and TDDFT, and \texttt{BerkeleyGW} for GW.
This section has been split on the basis of the calculation method/routine where both the accuracy and the performance of the code is discussed.
\begin{figure}[h]
    \centering
    \includegraphics[width=\textwidth]{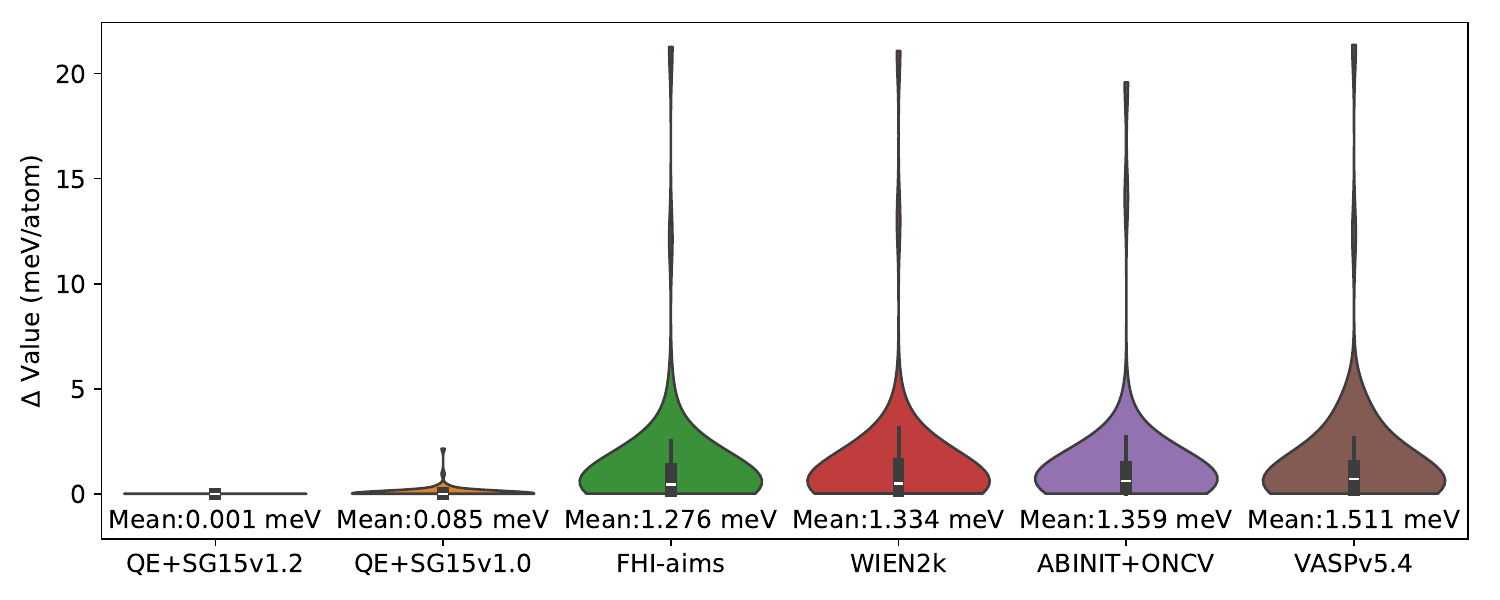}
    \caption{Distribution of element-wise $\Delta$ values computed between \texttt{Quantum MASALA} and the following DFT codes: \texttt{Quantum ESPRESSO} (with SG15 v1.2), \texttt{Quantum ESPRESSO} (with SG15 v1.0) \texttt{FHI-aims}, \texttt{WIEN2k}, \texttt{ABINIT} (with PseudoDojo ONCVPSP v0.1), and \texttt{VASPv5.4}}
    \label{fig2}
\end{figure}
\begin{table}[!htb]
    \centering
\begin{threeparttable}
\begin{tabular*}{\linewidth}{@{\hskip\tabcolsep\extracolsep{\fill}}
                c *{7}{S[table-format=2.2]}}
        \toprule
\multicolumn{1}{c}{}
    & \multicolumn{1}{c}{}
        & \multicolumn{2}{c}{\texttt{Quantum MASALA}}
            & \multicolumn{2}{c}{\texttt{BerkeleyGW}} 
                & \multicolumn{1}{c}{}\\
    \cmidrule(lr){3-4} \cmidrule(lr){5-6}
{$k$-pt.} & {LDA} & {Static} & {GPP} & {Static} & {GPP} &{Exp.}\\
    \midrule 
    $\Gamma_{1\textrm{v}}$   & -11.93 & -12.75 & -11.65 & -12.75 & -11.65 & -12.50$^{\text{a}}$ \\
    $\Gamma'_{25\textrm{v}}$ &   0.00 &   0.00 &   0.00 &   0.00 &   0.00 &   0.00$^{\text{a}}$ \\
    $\Gamma_{15\textrm{c}}$  &   2.56 &   3.87 &   3.36 &   3.87 &   3.36 &   3.34$^{\text{b}}$ \\
    $\Gamma'_{2\textrm{c}}$  &   3.26 &   4.32 &   3.94 &   4.32 &   3.94 &   4.18$^{\text{a}}$ \\[5pt]
    
    X$_{1\textrm{v}}$ & -7.78 & -8.31 & -7.75 & -8.31 & -7.75 &  ~ \\ 
    X$_{4\textrm{v}}$ & -2.86 & -2.86 & -2.85 & -2.86 & -2.85 & -2.90$^{\text{b}}$ \\
    X$_{1\textrm{c}}$ & 0.66  &  2.07 &  1.49 &  2.07 &  1.49 &  1.30$^{\text{c}}$ \\[5pt]
    
    L$'_{2\textrm{v}}$ & -9.58 & -10.28 & -9.47 & -10.28 & -9.47 & -9.30$^{\text{b}}$ \\ 
    L$_{1\textrm{v}}$  & -6.98 & -7.24  & -6.92 & -7.24  & -6.92 & -6.80$^{\text{b}}$ \\
    L$'_{3\textrm{v}}$ & -1.20 & -1.20  & -1.21 & -1.20  & -1.21 & -1.20$^{\text{b}}$ \\ 
    L$_{1\textrm{c}}$  &  1.49 &  2.64  &  2.22 &  2.64  &  2.22 &  2.04$^{\text{b}}$ \\ 
    L$_{3\textrm{c}}$  &  3.33 &  4.80  &  4.23 &  4.80  &  4.23 &  3.90$^{\text{b}}$ \\
    \bottomrule
\end{tabular*}
    \begin{tablenotes}[para, flushleft]\footnotesize
    \note   All values are in $eV$.\\
    \text{a}: Reference \cite{Chiang_Goldmann_1989} and references therein. \\
    \text{b}: Reference \cite{si_exp_1}. \\
    \text{c}: Reference \cite{Malone_2013}.
    \end{tablenotes}
\end{threeparttable}
\caption{Comparison of quasiparticle energies for Si, calculated using \texttt{Quantum MASALA} and \texttt{BerkeleyGW}. Experimental data taken from \cite{Malone_2013}, 
    \label{tab:GW_results}\cite{Chiang_Goldmann_1989}, and \cite{si_exp_1}.}
\end{table}
\subsection{DFT}
For DFT calculations, the $\Delta$ Benchmark\cite{DeltaBench} provides a standardized protocol that quantifies the differences in results obtained between different solid state codes.
The $\Delta$ value is the root-mean-square difference between the equation of state computed from two codes, averaged over a benchmark set of 71 elemental crystals.
The equation of state is determined by fitting the Birch-Murnaghan Equation of state\cite{BirchMurnaghan} using energies computed across different unit-cell volumes of the crystal.

The $\Delta$ Benchmark has been run with \texttt{Quantum MASALA} using the SG15\cite{SG15} set of ONCV pseudopotentials (ver 1.2). The generated Birch-Murnaghan parameters can be found in appendix \ref{appendix.delta} for reference. Figure (\ref{fig2}) shows the comparison of the results obtained from \texttt{Quantum MASALA} to several codes: \texttt{Quantum ESPRESSO} (with SG15 v1.2), \texttt{Quantum ESPRESSO} (with SG15 v1.0), \texttt{FHI-aims}, \texttt{WIEN2k}, \texttt{ABINIT} (with PseudoDojo ONCVPSP v0.1), and \texttt{VASPv5.4}. As can be seen from the figure, the results from \texttt{Quantum MASALA} agree excellently with those from other codes. The $\Delta$ errors between \texttt{Quantum MASALA} and these codes are comparable to the $\Delta$ errors between the codes. All electron DFT calculations are considered to be the standard as they are not affected by the pseudization of potential. The $\Delta$-value between \texttt{Quantum MASALA} and \texttt{WIEN2k}\cite{wien2k1,Weik2k2}, an all-electron DFT code is 1.3 meV/atom. We note that the results from \texttt{Quantum MASALA} match the results from \texttt{Quantum ESPRESSO}, using the same pseudopotentials to $\sim$1 $\mu$eV/atom. This demonstrates the accuracy of the DFT routines implemented in \texttt{Quantum MASALA}.

\subsection{GW}
For GWA calculations, we present a comparison of quasiparticle energy calculation results from \texttt{BerkeleyGW} and from \texttt{Quantum MASALA} for Static COHSEX, and Hybertsen-Louie Plasmon Pole with Static Remainder correction in table (\ref{tab:GW_results}). The convergence study presented in \cite{Malone_2013} was used to decide the parameters required for convergence of Silicon quasiparticle energies for $6\times6\times6$ k-grid within 10 meV. For the mean field calculation (using DFT), the wave function was expanded in plane waves with energy upto 25 Ry. A dielectric cut off of 25 Ry was used. 274 empty bands were included in Coulomb-hole summation. The calculated quasiparticle energies obtained using \texttt{Quantum MASALA} were found to match the results obtained using \texttt{BerkeleyGW} to within 100 $\mu$eV.

\subsection{TDDFT}
\begin{figure}[th!]
    \centering
    \includegraphics[width=0.8\textwidth]{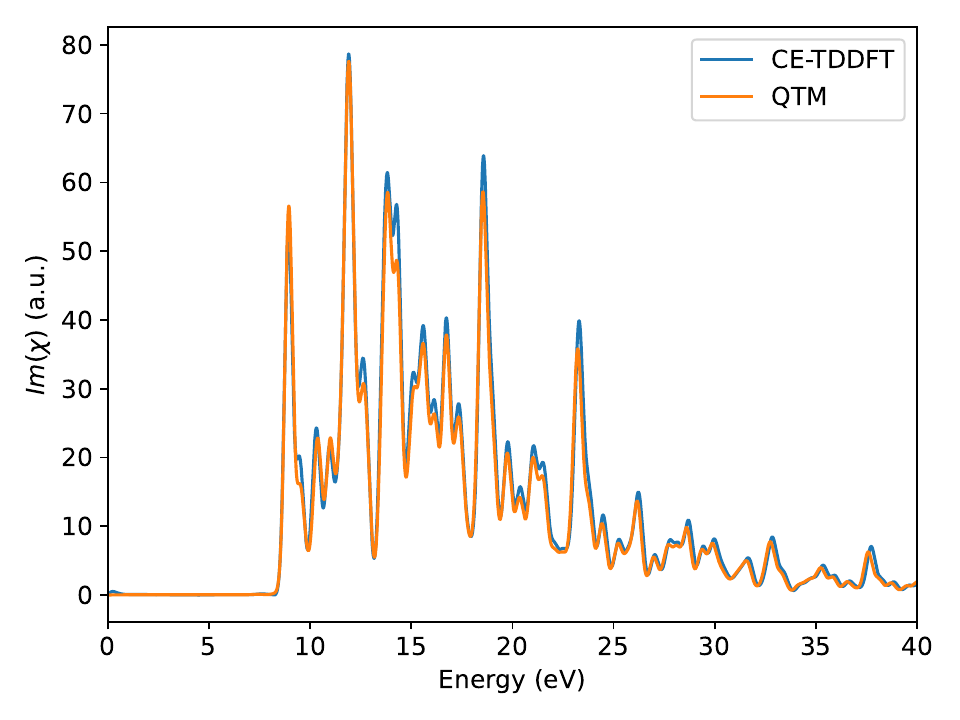}
    \caption{Absorption spectrum of Methane molecule obtained from \texttt{Quantum MASALA}'s TDDPT module (orange) and the TDDFT implementation in \texttt{CE-TDDFT} (blue)}
    \label{Fig:TDDFT}
\end{figure}

The accuracy of the TDDFT routines in \texttt{Quantum MASALA} is assessed by evaluating the dynamical polarizability matrix elements $\chi_{ij}(\omega)$, whose trace is proportional to the optical absorption spectrum $\sigma(\omega)$. This is checked against the same calculation performed using \texttt{CE-TDDFT}\cite{cetddft}. We have used a cubic box of side 30 Bohr, a wavefunction cutoff of 25 Rydbergs, and SG15\cite{SG15} set of ONCV pseudopotentials (ver 1.2) for the C and the H atoms. The C-H bond length was set to 1.114 {\AA}, and the time propagation was performed for 10,000 time steps of roughly 2.4 attoseconds each.
In figure (\ref{Fig:TDDFT}) we present the results obtained for the Methane molecule. The figure shows a comparison between the dynamical polarizability between \texttt{Quantum MASALA} and \texttt{CE-TDDFT}. From the figure, it is clear that the results agree well with each other. We plan to implement improved algorithms to enable the package to target bigger systems in the future.

\subsection{Performance}
Although \texttt{Quantum MASALA} is primarily designed to be compact and simple, the performance-critical sections of the code have been extensively optimized through efficient utilization of linear algebra routines provided by NumPy+Scipy(CPU)/CuPy(GPU). This allows the code to minimize the performance gap to its compiled counterparts. An ideal implementation is expected to maximize the fraction of runtime spent in library routines instead of the Python interpreter.
\begin{figure}[th!]
    \centering
    \includegraphics[width=0.8\textwidth]{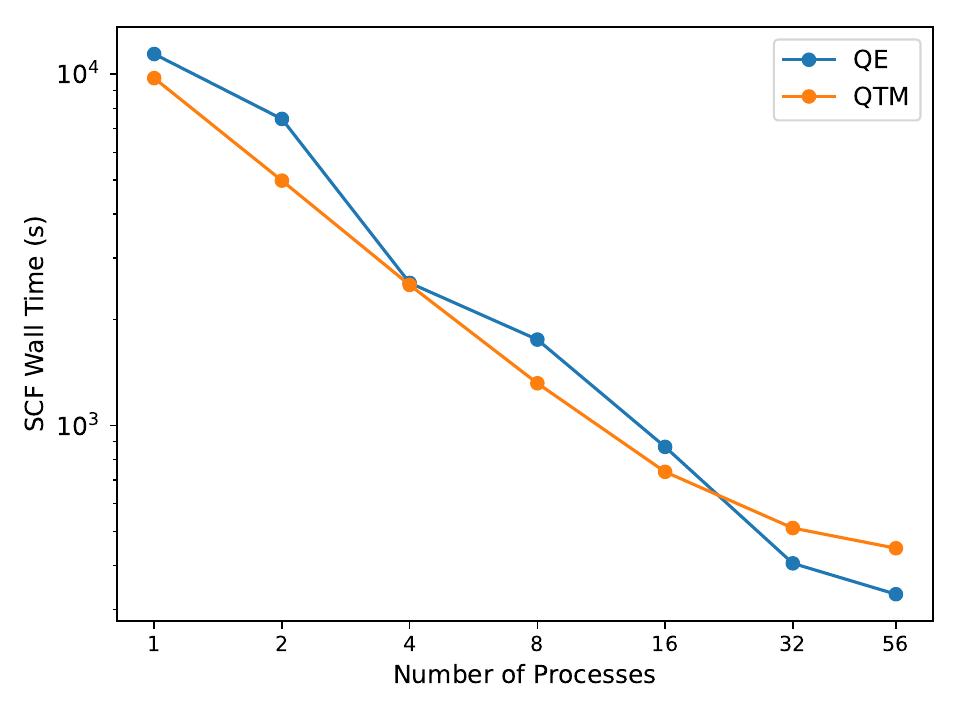}
    \caption{Wall time vs \# of processes used for SCF calculation of the Fe crystal in the $\Delta$ Benchmark with k-point parallelization only. Orange and blue points correspond to wall times for \texttt{Quantum MASALA} and \texttt{Quantum ESPRESSO} respectively.}
    \label{Fig:Fe}
\end{figure}

To assess the performance of \texttt{Quantum MASALA}, we have measured the run times of DFT calculations which were performed with varying number of processors.
The wall times for the SCF routine were compared against the same performed on an installation of \texttt{Quantum ESPRESSO} v7.2 that is linked to the same libraries/dependencies as \texttt{Quantum MASALA} (FFTW 3.3.10, Libxc 6.2.2, Intel MKL 2023.2.0, Intel MPI Library 2021.10.0).
The benchmarks were performed in a workstation equipped with two Intel(R) Xeon(R) Gold 6348 processors and around 1 Terabyte of DDR4 3200 MHz RAM. It must be noted that multi-threading is disabled in our benchmarks. 

\begin{figure}[th!]
    \centering
    \includegraphics[width=0.8\textwidth]{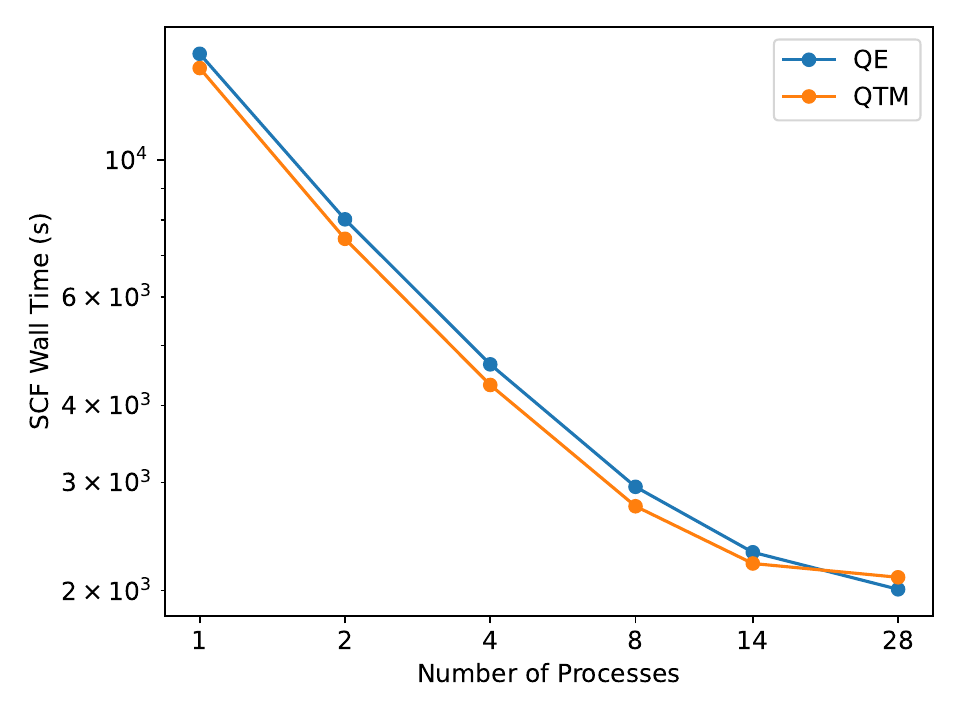}
    \caption{Wall time vs \# of processes used for SCF calculation of a $6\times 6\times 6$ supercell of silicon ($\Gamma$-point only) with parallelization across bands. Orange and blue points correspond to wall times for \texttt{Quantum MASALA} and \texttt{Quantum ESPRESSO} respectively.}
    \label{Fig:Si_bands}
\end{figure}

\begin{figure}[th!]
    \centering
    \includegraphics[width=0.8\textwidth]{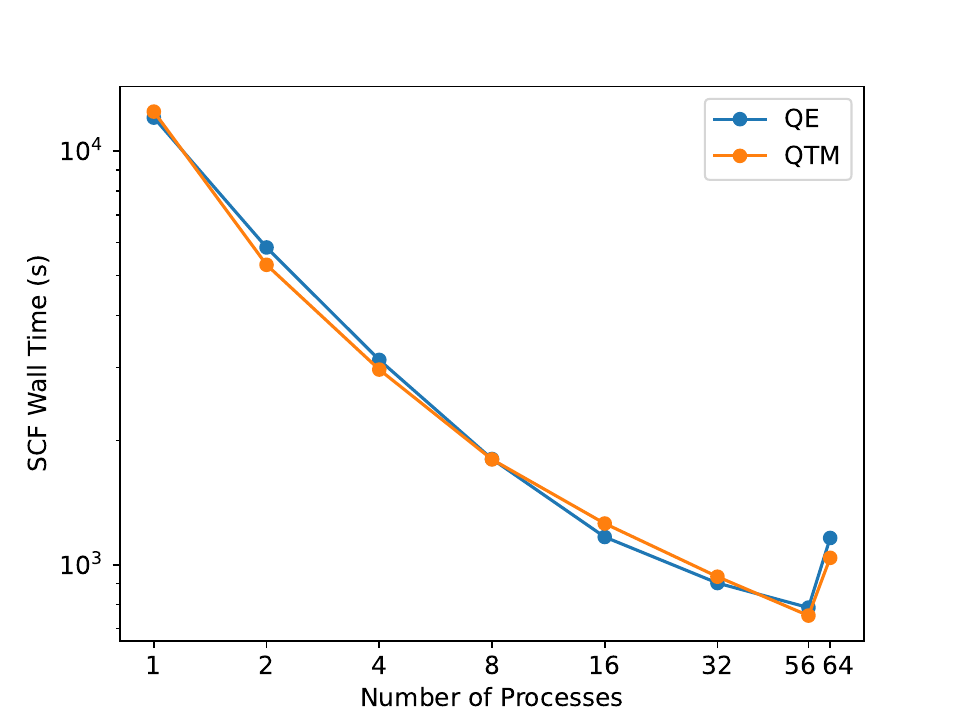}
    \caption{Wall time vs \# of processes used for SCF calculation of a $6\times 6\times 6$ supercell of silicon ($\Gamma$-point only) with parallelization across G-space. Orange and blue points correspond to wall times for \texttt{Quantum MASALA} and \texttt{Quantum ESPRESSO} respectively.}
    \label{Fig:Si_gspace}
\end{figure}

We present results from three benchmarks, each targeting a particular parallelization mode. The first, a crystalline Iron system used in delta benchmark. In this benchmark, the k-points for sampling the Brillouin zone are distributed among MPI tasks. Figure (\ref{Fig:Fe}) shows the reduction in the wall time taken as the number of processors (same as MPI tasks) are increased keeping the number of k-points the same. It also shows the timings for the same calculation using the k-point parallelization in \texttt{Quantum ESPRESSO}. The details of the calculation and exact times for this benchmark are shown in appendix \ref{appendix.kpoint_bench}. The second is the band parallelization benchmark. This benchmark is done on a Silicon supercell containing 6$\times$6$\times$6 FCC primitive unit-cells (432 atoms in total), with only the $\Gamma$ point sampling of the Brillouin zone. Figure (\ref{Fig:Si_bands}) shows the reduction in the wall time taken as the number of processors are increased keeping the number of bands constant. The figure also shows the scaling for \texttt{Quantum ESPRESSO} on the same problem. The details of the calculation are given in appendix \ref{appendix.bpar_bench}. In the third benchmark, we have used the same Silicon supercell system (as the second benchmark) to study plane-wave parallelization. We note, that while the bands or G-space elements are distributed among various MPI tasks, the final diagonalization of the subspace projected matrix is performed serially in both the codes for the presented benchmarks. Figure(\ref{Fig:Si_gspace}) shows the wall time as a function of MPI tasks keeping the number of G-vectors constant. We also compare the wall times to \texttt{Quantum ESPRESSO} for the same calculation using identical parallelization scheme (and serial diagonalization). The exact wall time numbers and parameters for this performance benchmark are presented in appendix \ref{appendix.gspace_bench}.
Figures (\ref{Fig:Fe}), (\ref{Fig:Si_bands}) and (\ref{Fig:Si_gspace}) show that \texttt{Quantum MASALA} can tackle medium-scale systems with performance comparable to that of \texttt{Quantum ESPRESSO}. We note that we were able to run a calculation with 1024 Si atoms (and parameters similar to the G-space/band parallelization benchmark) on a single node (with configuration mentioned above) using \texttt{Quantum MASALA}. This calculation took about 3 hours to finish. 

On a 2/7 partition of an NVidia A100 80GB card, SCF calculation for the same Silicon 6$\times$6$\times$6 supercell system runs in 332 seconds. See appendix \ref{appendix.gpu_runs} for more details. We note that this time is about a factor of 2 faster than the fastest time obtained using G-space parallelization on the same problem. 

The above mentioned results clearly show that \texttt{Quantum MASALA} is not only efficient in terms of number of lines of code but also in terms of both serial and parallel performance. This enables implementations built using this code to not be bottle-necked by the relatively slow programming language while leveraging its simplicity for rapid prototyping.

\section{Conclusion}
We have presented \texttt{Quantum MASALA}, a compact package that implements different electronic-structure methods in Python.
Within just 8100 lines of pure Python code, we have implemented Density Functional Theory (DFT), Time-dependent Density Functional Theory (TD-DFT) and the GW Method for periodic systems.
We have demonstrated the ability of the code to scale across multiple process cores and to run in Graphical Processing Units (GPU) with the help of easily-accessible Python libraries.
With \texttt{Quantum ESPRESSO} and \texttt{BerkeleyGW} input interfaces implemented, it can also be used as a substitute for small and medium scale calculations. \texttt{Quantum MASALA} is extremely light and simple, making it an ideal tool for teaching and learning {\it ab initio} methods. Further, its performance and modularity makes it a perfect framework for developing and testing new methods for {\it ab initio} electronic structure calculations.

\section*{Acknowledgements}
S.H. thanks Department of Science and Technology (DST), Govt. of India for financial support through Kishore Vaigyanik Protsahan Yojana (KVPY) fellowship.
A.S. thanks DST, Govt. of India for financial support through the KVPY and the Prime Minister's Reseach Fellowship (PMRF) fellowships and University Grants Commisssion, Govt. of India for financial support through the UGC-JRF fellowship.
This work was completed in part at the GPU Application Hackathon 2023, part of the Open Hackathons program. The authors would like to acknowledge OpenACC-Standard.org, NVidia and the National Supercomputing Mission for their support, and Prof. Shivasubramanian Gopalakrishnan, and Dr. Mukul Dave for their mentorship during the event. The authors would like to thank Prof. Rahul Pandit for access to the NVidia A100 GPU. 
The authors would also like to thank Prof. H. R. Krishnamurthy, Varshitha K. S., Vaibhav Sahu, Shinjan Mandal, Ishita Shitut, Namana Venkatareddy, Robin Bajaj, Harsimran Singh, Sanmit Chakraborty, and Victor Ghosh for useful discussions and comments during the development of the code.




\appendix
\section*{Appendix}
\addcontentsline{toc}{section}{Appendices}
\renewcommand{\thesubsection}{\Alph{subsection}}

\subsection{$\Delta$ Benchmark: Complete Results}\label{appendix.delta}
The results for the $\Delta$ benchmark run with parameters same as Ref. \cite{DeltaBench}. We used SG15 version 2 pseudopotentials \cite{SG15}. Table (\ref{tab:delta_bench}) shows the results for each element in the benchmark.
\addtolength{\tabcolsep}{-3pt}
\captionof{table}{Result of the $\Delta$ Benchmark of \texttt{Quantum MASALA}: Equation of State (EOS) and $\Delta$-value with respect to \texttt{WIEN2k}}
\label{tab:delta_bench}
\begin{tabularx}{\textwidth}{c|cc|YYY|YYY}
\multicolumn{1}{c|}{} & \multicolumn{1}{c}{$Z_{val}$} & \multicolumn{1}{c|}{k-mesh}& \multicolumn{1}{c}{$V_0$} & \multicolumn{1}{c}{$B_0$} & \multicolumn{1}{c|}{$B_1$} & \multicolumn{1}{c}{$\Delta_{\textrm{WIEN2k}}$} & \multicolumn{1}{c}{$\Delta_{\textrm{QE,1.0}}$}& \multicolumn{1}{c}{$\Delta_{\textrm{QE,1.2}}$}  \\
 & & &\multicolumn{1}{c}{[\AA/at]}&\multicolumn{1}{c}{[GPa]}&\multicolumn{1}{c|}{[-]}&\multicolumn{1}{c}{[meV/at]}&\multicolumn{1}{c}{[meV/at]}& \multicolumn{1}{c}{[meV/at]}\\
 \hline
 \endhead
H  & 1   & 28×28×20                &  17.356  &   10.287 &  2.670               & 0.072	 & 0.000                    & 0.000                   \\
He & 2   & 40×40×22                &  17.709  &    0.881 &  6.454               & 0.013	 & 0.000                    & 0.000                   \\
Li & 3   & 38×38×38                &  20.243  &   13.846 &  3.347               & 0.074	 & 0.000                    & 0.000                   \\
Be & 4   & 52×52×28                &   7.932  &  123.695 &  3.308               & 0.610	 & 0.009                    & 0.008                   \\
B  & 3   & 26×26×24                &   7.180  &  235.365 &  3.427               & 3.129	 & 0.003                    & 0.001                   \\
C  & 4   & 48×48×12                &  11.588  &  207.052 &  3.589               & 2.223	 & 0.052                    & 0.000                   \\
N  & 5   & 16×16×16                &  28.773  &   53.219 &  3.662               & 1.353	 & 0.001                    & 0.006                   \\
O  & 6   & 26×24×24                &  18.597  &   50.520 &  3.919               & 0.419	 & 0.054                    & 0.000                   \\
F  & 7   & 16×28×14                &  19.284  &   33.756 &  3.995               & 0.867	 & 0.002                    & 0.000                   \\
Ne & 8   & 22×22×22                &  24.259  &    1.352 &  7.251               & 0.021	 & 0.000                    & 0.000                   \\
Na & 9   & 32×32×32                &  37.113  &    7.762 &  3.698               & 0.587	 & 0.000                    & 0.000                   \\
Mg & 10  & 36×36×20                &  22.961  &   36.538 &  4.049               & 0.222	 & 0.002                    & 0.000                   \\
Al & 11  & 24×24×24                &  16.528  &   77.780 &  4.932               & 0.861	 & 0.005                    & 0.000                   \\
Si & 4   & 32×32×32                &  20.543  &   87.434 &  4.265               & 1.701	 & 0.387                    & 0.001                   \\
P  & 5   & 30×8×22                 &  21.475  &   67.802 &  4.311               & 0.064	 & 0.028                    & 0.000                   \\
S  & 6   & 38×38×38                &  17.293  &   83.977 &  4.065               & 1.988	 & 0.945                    & 0.000                   \\
Cl & 7   & 12×24×12                &  39.328  &   18.707 &  4.377               & 1.806	 & 0.236                    & 0.000                   \\
Ar & 8   & 16×16×16                &  52.482  &    0.752 &  7.759               & 0.019	 & 0.027                    & 0.000                   \\
K  & 9   & 20×20×20                &  73.654  &    3.602 &  4.003               & 0.032	 & 0.001                    & 0.000                   \\
Ca & 10  & 18×18×18                &  42.172  &   17.634 &  3.413               & 0.137	 & 0.002                    & 0.000                   \\
Sc & 11  & 34×34×20                &  24.643  &   54.546 &  3.383               & 0.271	 & 0.003                    & 0.000                   \\
Ti & 12  & 40×40×22                &  17.407  &  112.062 &  3.541               & 0.407	 & 0.001                    & 0.000                   \\
V  & 13  & 34×34×34                &  13.468  &  182.313 &  3.962               & 0.692	 & 0.006                    & 0.002                   \\
Cr & 14  & 36×36×36                &  12.441  &  115.340 &  6.981               & 21.077 & 0.010                    & 0.002                   \\
Mn & 15  & 28×28×28                &  11.899  &  128.145 &  4.529               & 12.994 & 0.062                    & 0.006                   \\
Fe & 16  & 36×36×36                &  11.454  &  180.301 &  6.856               & 4.635	 & 0.072                    & 0.004                   \\
Co & 17  & 46×46×24                &  10.923  &  211.257 &  4.764               & 2.986	 & 0.012                    & 0.000                   \\
Ni & 18  & 28×28×28                &  10.943  &  195.432 &  5.075               & 2.366	 & 0.003                    & 0.000                   \\
Cu & 19  & 28×28×28                &  11.986  &  138.902 &  5.104               & 1.099	 & 0.003                    & 0.000                   \\
Zn & 20  & 44×44×20                &  15.172  &   75.758 &  5.350               & 0.155	 & 0.004                    & 0.001                   \\
Ga & 13  & 22×12×22                &  20.339  &   49.771 &  6.242               & 0.438	 & 0.142                    & 0.000                   \\
Ge & 14  & 30×30×30                &  23.981  &   58.976 &  4.890               & 0.847	 & 0.002                    & 0.000                   \\
As & 5   & 30×30×10                &  22.673  &   68.072 &  4.253               & 1.258	 & 0.235                    & 0.000                   \\
Se & 6   & 26×26×20                &  29.806  &   46.930 &  4.433               & 0.638	 & 2.146                    & 0.005                   \\
Br & 7   & 12×24×12                &  39.670  &   22.378 &  4.832               & 1.091	 & 0.000                    & 0.000                   \\
Kr & 8   & 16×16×16                &  65.982  &    0.647 &  7.300               & 0.037	 & 0.000                    & 0.000                   \\
Rb & 9   & 18×18×18                &  91.029  &    2.794 &  3.721               & 0.095	 & 0.005                    & 0.000                   \\
Sr & 10  & 16×16×16                &  54.399  &   11.252 &  5.392               & 0.226	 & 0.013                    & 0.000                   \\
Y  & 11  & 32×32×18                &  32.858  &   41.389 &  3.156               & 0.138	 & 0.001                    & 0.000                   \\
Zr & 12  & 36×36×20                &  23.398  &   94.022 &  3.282               & 0.282	 & 0.001                    & 0.000                   \\
Nb & 13  & 30×30×30                &  18.148  &  169.942 &  3.602               & 0.444	 & 0.017                    & 0.000                   \\
Mo & 14  & 32×32×32                &  15.781  &  260.286 &  4.316               & 0.277	 & 0.008                    & 0.001                   \\
Tc & N/A & N/A                     &   \multicolumn{1}{r}{N/A}  &      \multicolumn{1}{r}{N/A} &    \multicolumn{1}{r|}{N/A}               & \multicolumn{1}{r}{N/A} 	  & \multicolumn{1}{r}{N/A}                     & \multicolumn{1}{r}{N/A} 	                   \\
Ru & 16  & 42×42×24                &  13.770  &  312.146 &  4.848               & 0.491	 & 0.002                    & 0.002                   \\
Rh & 17  & 26×26×26                &  14.050  &  256.931 &  5.168               & 0.558	 & 0.010                    & 0.001                   \\
Pd & 18  & 26×26×26                &  15.311  &  169.638 &  5.547               & 0.089	 & 0.009                    & 0.001                   \\
Ag & 19  & 24×24×24                &  17.825  &   91.057 &  6.044               & 0.351	 & 0.003                    & 0.002                   \\
Cd & 20  & 38×38×18                &  22.950  &   43.615 &  6.965               & 3.136	 & 0.001                    & 0.000                   \\
In & 13  & 30×30×20                &  27.530  &   36.075 &  4.986               & 0.528	 & 0.119                    & 0.000                   \\
Sn & 14  & 26×26×26                &  36.836  &   35.730 &  4.911               & 0.176	 & 0.331                    & 0.000                   \\
Sb & 15  & 26×26×8                 &  31.773  &   50.379 &  4.536               & 0.482	 & 0.121                    & 0.000                   \\
Te & 16  & 26×26×16                &  35.075  &   44.639 &  4.616               & 0.954	 & 0.377                    & 0.001                   \\
I  & 17  & 12×22×10                &  50.398  &   18.595 &  5.046               & 0.670	 & 0.093                    & 0.000                   \\
Xe & 18  & 14×14×14                &  87.128  &    0.538 &  7.168               & 0.056	 & 0.040                    & 0.000                   \\
Cs & 9   & 16×16×16                & 116.870  &    1.969 &  3.907               & 0.061	 & 0.010                    & 0.000                   \\
Ba & 10  & 20×20×20                &  63.229  &    8.728 &  1.994               & 0.098	 & 0.004                    & 0.000                   \\
Lu & N/A & N/A                     &     \multicolumn{1}{r}{N/A}  &      \multicolumn{1}{r}{N/A} &    \multicolumn{1}{r|}{N/A}               & \multicolumn{1}{r}{N/A} 	  & \multicolumn{1}{r}{N/A}                     & \multicolumn{1}{r}{N/A} 	                   \\
Hf & 26  & 36×36×20                &  22.596  &  108.624 &  3.442               & 1.512	 & 0.001                    & 0.000                   \\
Ta & 27  & 30×30×30                &  18.325  &  195.050 &  3.615               & 1.631	 & 0.005                    & 0.000                   \\
W  & 28  & 32×32×32                &  16.150  &  301.095 &  4.130               & 0.638	 & 0.003                    & 0.001                   \\
Re & 15  & 42×42×22                &  14.940  &  365.166 &  4.405               & 1.454	 & 0.021                    & 0.001                   \\
Os & 16  & 42×42×24                &  14.255  &  398.968 &  4.809               & 2.195	 & 0.014                    & 0.001                   \\
Ir & 17  & 26×26×26                &  14.472  &  349.246 &  5.115               & 2.149	 & 0.006                    & 0.001                   \\
Pt & 18  & 26×26×26                &  15.687  &  246.047 &  5.473               & 2.413	 & 0.003                    & 0.006                   \\
Au & 19  & 24×24×24                &  17.982  &  138.727 &  6.014               & 0.278	 & 0.002                    & 0.001                   \\
Hg & 20  & 24×24×28                &  29.550  &    7.795 &  9.984               & 0.078	 & 0.004                    & 0.000                   \\
Tl & 13  & 32×32×18                &  31.360  &   26.876 &  5.468               & 0.177	 & 0.001                    & 0.000                   \\
Pb & 14  & 20×20×20                &  31.945  &   40.064 &  5.481               & 0.401	 & 0.013                    & 0.000                   \\
Bi & 15  & 26×26×8                 &  36.917  &   42.819 &  4.693               & 0.127	 & 0.002                    & 0.000                   \\
Po & N/A & N/A                     &     \multicolumn{1}{r}{N/A}  &      \multicolumn{1}{r}{N/A} &    \multicolumn{1}{r|}{N/A}               & \multicolumn{1}{r}{N/A}    & \multicolumn{1}{r}{N/A}                       & \multicolumn{1}{r}{N/A}                       \\
Rn & N/A & N/A                     &     \multicolumn{1}{r}{N/A}  &      \multicolumn{1}{r}{N/A} &    \multicolumn{1}{r|}{N/A}               & \multicolumn{1}{r}{N/A}    & \multicolumn{1}{r}{N/A}                       & \multicolumn{1}{r}{N/A}                       \\
\hline
& & & & & \multicolumn{1}{r|}{$\expval{\Delta}$} & 1.298 & 0.089 & 0.001\\
& & & & & \multicolumn{1}{r|}{$\sigma_{\Delta}$} & 2.980 & 0.289 & 0.002
\end{tabularx}

\subsection{Performance: Scaling under k-point parallelization}\label{appendix.kpoint_bench}

Table (\ref{kpoint_bench}) shows the scaling of wall time for SCF routines with a fixed number of k-points as the number of MPI processes are increased within the k-point parallelization scheme. These spin-polarized SCF calculations were performed on a BCC Iron unit cell. The Perdew-Burke-Ernzerhof (PBE) exchange-correlation functional was used for these calculation and the ONCV pseudopotential was taken from the SG15\cite{SG15} set (version 1.2). Kinetic energy cutoff for wavefunctions was set to 100 Ry, and a shifted Monkhorst-Pack grid of $36\times36\times36$ k-points was used. Broyden mixing was used for the densities, with a mixing factor of 0.3. The SCF convergence threshold was set to $10^{-10}$ Ry. Gaussian smearing was used with a fictitious temperature of 0.01 eV ($\sim 7.35\times10^{-4}$ Ry). The table clearly shows that the scaling of \texttt{Quantum MASALA} compares well with \texttt{Quantum ESPRESSO}.

\begin{table}[!h]
    \centering
    \begin{tabular}{ccccc}
        \toprule
         Number of & \multicolumn{2}{c}{SCF Wall Time (s)} & \multicolumn{2}{c}{Number of H-psi calls} \\ 
         \cmidrule(lr){2-3} \cmidrule(lr){4-5}
        Processes & QE & QTM & QE & QTM \\ 
        \midrule
        1 & 11415 & 9748 & 767985 & 692283 \\ 
        2 & 7459 & 4978 & 503570 & 337008 \\ 
        4 & 2547 & 2513 & 168153 & 168275 \\ 
        8 & 1757 & 1319 & 105430 & 84157 \\ 
        16 & 869 & 738 & 53923 & 43589 \\ 
        32 & 405 & 511 & 22061 & 20996 \\ 
        56 & 331 & 447 & 13721 & 12169 \\ 
        \bottomrule
    \end{tabular}
    \caption{Scaling of Wall time for SCF routines of \texttt{Quantum ESPRESSO} (QE) and \texttt{Quantum MASALA} (QTM) versus the number of MPI processes under k-point parallelization only. The SCF calculation was performed on Fe unitcell.}
    \label{kpoint_bench}
\end{table}

\subsection{Performance: Scaling under band parallelization}\label{appendix.bpar_bench}

Table (\ref{bpar_bench}) shows the scaling of wall time for SCF routines with a fixed number of bands as the number of MPI processes are increased within the band parallelization scheme. These non-spin-polarized SCF calculations were performed on a $6\times6\times6$ Silicon supercell, containing a total of 432 silicon atoms, using only band parallelization. We used the Perdew-Burke-Ernzerhof (PBE) exchange-correlation functional and an ONCV pseudopotential available from the SG15\cite{SG15} set (version 1.2). Kinetic energy cutoff for wavefunctions was set to 25 Ry, and $\Gamma$-point calculations were performed. Broyden mixing was used for the densities, with a mixing factor of 0.7. The SCF convergence threshold was set to $10^{-10}$ Ry. The table clearly shows that the scaling of \texttt{Quantum MASALA} compares well with \texttt{Quantum ESPRESSO}.

\begin{table}[!h]
    \centering
    \begin{tabular}{ccccc}
        \toprule
         Number of & \multicolumn{2}{c}{SCF Wall Time (s)} & \multicolumn{2}{c}{Number of H-psi calls} \\ 
         \cmidrule(lr){2-3} \cmidrule(lr){4-5}
        Processes & QE & QTM & QE & QTM \\ 
        \midrule
        1 & 14886 & 14112 & 158 & 119 \\ 
        2 & 8017 & 7454 & 158 & 119 \\ 
        4 & 4662 & 4316 & 158 & 119 \\ 
        8 & 2949 & 2743 & 158 & 119 \\ 
        14 & 2309 & 2214 & 158 & 119 \\ 
        28 & 2011 & 2103 & 158 & 117 \\ 
        \bottomrule
    \end{tabular}
    \caption{Scaling of Wall time for SCF routines of \texttt{Quantum ESPRESSO} (QE) and \texttt{Quantum MASALA} (QTM) versus the number of MPI processes under band parallelization only. The SCF calculation was performed on Si $6\times6\times6$ supercell.}
    \label{bpar_bench}
\end{table}

\subsection{Performance: Scaling under G-Space parallelization}\label{appendix.gspace_bench}

Table (\ref{gspace_bench}) shows the scaling of wall time for SCF routines for a fixed G-space size as the number of MPI processes are increased within the G-space parallelization scheme. For these set of calculations, we used the same system and calculation parameters as in appendix \ref{appendix.bpar_bench}, with the exception that the SCF convergence threshold was set to $10^{-8}$ Ry. The table clearly shows that the scaling of \texttt{Quantum MASALA} compares well with \texttt{Quantum ESPRESSO}.

\begin{table}[!h]
    \centering
    \begin{tabular}{ccccc}
        \toprule
         Number of & \multicolumn{2}{c}{SCF Wall Time (s)} & \multicolumn{2}{c}{Number of H-psi calls} \\ 
         \cmidrule(lr){2-3} \cmidrule(lr){4-5}
        Processes & QE & QTM & QE & QTM \\ 
        \midrule        
        1 & 12040 & 12449 & 110 & 107 \\ 
        2 & 5845 & 5307 & 102 & 97 \\ 
        4 & 3126 & 2962 & 98 & 98 \\ 
        8 & 1801 & 1797 & 103 & 101 \\ 
        16 & 1167 & 1258 & 96 & 101 \\ 
        32 & 903 & 935 & 93 & 98 \\ 
        56 & 787 & 753 & 98 & 95 \\ 
        64 & 1160 & 1039 & 102 & 92 \\ 
        \bottomrule
    \end{tabular}
    \caption{Scaling of Wall time for SCF routines of \texttt{Quantum ESPRESSO} (QE) and \texttt{Quantum MASALA} (QTM) versus the number of MPI processes under G-space parallelization only. The SCF calculation was performed on Si $6\times6\times6$ supercell.}
    \label{gspace_bench}
\end{table}

\subsection{Performance: Scaling on GPU}\label{appendix.gpu_runs}

Table (\ref{gpu_scaling}) shows the scaling of wall time for SCF routines for increasing problem size within the GPU. For these set of calculations, we used the same parameters as in appendix \ref{appendix.gspace_bench}, and ran $\Gamma$-point SCF calculations for Silicon, with varying supercell sizes. We ran the calculations on a 2/7 partition of an NVidia A100 80GB GPU. Comparing the fastest time for the same supercell size from table (\ref{gpu_scaling}) and table (\ref{gspace_bench}), one can see that \texttt{Quantum MASALA} can deliver significant speedup ($\sim$ \textrm{factor of} 2) when using a GPU.

\begin{table}[!h]
    \centering
    \begin{tabular}{cYYYc}
        \toprule
        Supercell size &  \multicolumn{3}{c}{Wall Time (s)} & Number of \\
        \cmidrule(lr){2-4}
         &  \multicolumn{1}{c}{SCF} &  \multicolumn{1}{c}{H-psi} & \multicolumn{1}{c}{Diagonalization} & H-psi calls \\ 
        \midrule
            $1\times1\times1$ &      4.50 &       0.05 &            0.08 &            44 \\
            $2\times2\times2$ &      6.08 &       0.38 &            0.22 &            83 \\
            $3\times3\times3$ &     10.63 &       1.27 &            0.53 &            71 \\
            $4\times4\times4$ &     32.56 &       6.02 &            7.06 &            87 \\
            $5\times5\times5$ &    101.13 &      28.06 &           29.62 &            99 \\
            $6\times6\times6$&    322.15 &      93.50 &          117.81 &            93 \\
        \bottomrule
    \end{tabular}
    \caption{Scaling of Wall time for SCF routines of \texttt{Quantum MASALA} (QTM) versus  supercell size.}
    \label{gpu_scaling}
\end{table}



\bibliography{refs-theory.bib, refs-codes.bib,refs-py.bib}

\nolinenumbers

\end{document}